\documentclass[12pt]{article}
\pdfoutput=1
\usepackage[top=30truemm,bottom=30truemm,left=25truemm,right=25truemm]{geometry}
\usepackage{authblk}
\usepackage{amsthm}
\usepackage{amsmath,amssymb,bm}
\usepackage{cite}
\usepackage[]{hyperref}
\hypersetup{colorlinks,linkcolor={blue},citecolor={blue},urlcolor={blue}}  
\usepackage[titletoc]{appendix}
\usepackage{url}
\usepackage{graphicx,color}
\usepackage{bbold}
\usepackage{braket}
\usepackage{qcircuit}
\usepackage{here}
\usepackage{mathrsfs}

\newtheorem*{definition*}{Definition}
\newtheorem*{assumption*}{Assumption}

\newtheorem{innercustomthm}{Theorem}
\newenvironment{customthm}[1]
  {\renewcommand\theinnercustomthm{#1}\innercustomthm}
  {\endinnercustomthm}

\newtheorem{innercustomprop}{Proposition}
\newenvironment{customprop}[1]
  {\renewcommand\theinnercustomprop{#1}\innercustomprop}
  {\endinnercustomprop}
  
\newtheorem{innercustomcoro}{Corollary}
\newenvironment{customcoro}[1]
  {\renewcommand\theinnercustomcoro{#1}\innercustomcoro}
  {\endinnercustomcoro}
  
\newtheorem{innercustomconj}{Conjecture}
\newenvironment{customconj}[1]
  {\renewcommand\theinnercustomconj{#1}\innercustomconj}
  {\endinnercustomconj}

 \def\CD{{\cal D}}
 \def\CE{{\cal E}}

 \def\CH{{\cal H}}

\def\be{\begin{equation}}
\def\ee{\end{equation}}
\def\ba{\begin{eqnarray}}
\def\ea{\end{eqnarray}}

\numberwithin{equation}{section}

\AtBeginDocument{%
  {\footnotesize\global\footnotesep=0.8\baselineskip}%
}

\begin{document}

\begin{titlepage}
\thispagestyle{empty}

\begin{flushright}
YITP-22-138

\end{flushright}

\bigskip

\begin{center}
\noindent{\bf \LARGE Counting atypical black hole microstates }\\
\vspace{0.5cm}
\noindent{\bf \LARGE from entanglement wedges}\\
%\vspace{1.6cm}
\vspace{1.4cm}

{\bf \normalsize Zixia Wei$^{a,b}$ and Yasushi Yoneta$^{c}$}
\vspace{1cm}\\

${}^a${\it
Center for Gravitational Physics and Quantum Information,\\[-1mm]
Yukawa Institute for Theoretical Physics,
Kyoto University, Kyoto 606-8502, Japan
}\\[1.5mm]

${}^b${\it
Kavli Institute for Theoretical Physics, \\[-1mm]
University of California, Santa Barbara, CA 93106, USA
}\\[1.5mm]

% ${}^c${\it
% Komaba Institute for Science, \\[-1mm]
% The University of Tokyo,
% 3-8-1 Komaba, Meguro, Tokyo 153-8902, Japan
% }\\[1.5mm]

${}^c${\it
Department of Basic Science, \\[-1mm]
The University of Tokyo,
3-8-1 Komaba, Meguro, Tokyo 153-8902, Japan
}

\vskip 3em
\end{center}

\begin{abstract}
Disentangled black hole microstates are atypical states in holographic CFTs whose gravity duals do not have smooth horizons. 
If there exist sufficiently many disentangled microstates to account for the entire black hole entropy, then any black hole microstate can be written as a superposition of states without smooth horizons. 
We show that there exist sufficiently many disentangled microstates to account for almost the entire black hole entropy of a large AdS black hole at the semiclassical limit $G_N\rightarrow 0$.
In addition, we also argue that in generic quantum many-body systems with short-ranged interactions, there exist sufficiently many area law states in the microcanonical subspace to account for almost the entire thermodynamic entropy in the standard thermodynamic limit.
Area law states are atypical since a typical state should contain volume law entanglement.
Furthermore, we also present an explicit way
to construct such a set of area law states, and argue that the same construction may also be used to construct disentangled states.
% in the microcanonical subspace.
% to construct $\exp\left[S(1-o(1))\right]$ orthogonal area law states in the corresponding microcanonical subspace

%Assuming there is a well-defined semiclassical geometry for a disentangled microstate, it must have a nontrivial structure nearby the black hole horizon. In a similar way, we also show that 

% The proof follows by first considering the microcanonical ensemble in a holographic CFT, and then giving an upper bound of its entanglement of formation with holographic entanglement of purification. In a similar way, we also show that in general quantum many-body systems with local interactions, in the thermodynamic limit, the sub-Hilbert space $\CH_{[E-\Delta E,E]}$ expanded by the energy eigenstates in the energy shell $[E-\Delta E,E]$ has an orthogonal basis in which $O(e^{S/3})$ states are atypical in the sense that they only contain area law entanglement. By contrast, a typical state should contain volume law entanglement. This is shown by upper bounding entanglement of formation with reflected entropy, which is looser than entanglement of purification, but evaluable in general systems. Besides the proof of existence, we also present an algorithm to construct such a basis. 

\end{abstract}

\end{titlepage}

\newpage
\setcounter{page}{1}
\tableofcontents

\newpage

\section{Introduction}
The notion of typicality \cite{Goldstein2006,Popescu2006,Reimann2007,Muller2015} lies in the most fundamental part of statistical mechanics,
which provides an explanation for the basic principles of equilibrium statistical mechanics \cite{Mori2018}.
% which allows us to use various probability ensembles to equivalently predict the macroscopic features of quantum many-body systems.
On the other hand, generic or collective features of atypical states are much less investigated.
One reason for the lack of interest is because generic atypical states in generic quantum many-body systems will eventually thermalize and become a typical state under time evolution, and hence are difficult to be related to experimental observations. 

However, studying atypical states appear to be useful in a seemingly totally different context: black hole physics. General relativity predicts the existence of a smooth horizon as well as a black hole interior, which leads to mysteries such as the information paradox\cite{Hawking75,Page93} and the AMPS firewall paradox \cite{AMPS12,AMPSS13} when combined with fundamental principles in quantum theory. A long-standing yet crucial question in the study of quantum gravity is whether microstates of black holes also have smooth horizons, or the existence of smooth horizons and black hole interior is merely an emergent illusion of (semi-)classical gravity after coarse-graining from a UV complete quantum gravitational theory \cite{Mathur05,AMPS12,AMPSS13,NVW12,HM13}. 

In this context, it is enlightening to consider a class of atypical states called disentangled states \cite{HP20} in holographic conformal field theories. These are states which have a large entanglement deficit compared to typical states. According to the AdS/CFT correspondence \cite{Maldacena97} and the holographic entanglement entropy formulae \cite{RT06,RT06b,HRT07,Wall12,FLM13,EW15}, a disentangled state on the CFT side cannot be dual to AdS geometry which has a smooth horizon \cite{HP20}.\footnote{Note that, however, this is a little bit subtle since the concept of horizon is often associated with dynamical features, and a disentangled state may turn into a typical state under time evolution. Therefore, we just regard this as a possible interpretation of disentangled states. Validity of the results in this paper does not rely on the validity of this interpretation.} In other words, a disentangled state either corresponds to a geometry that has nontrivial structures at or outside the horizon, or does not have a geometrical description at all \cite{HP20}.\footnote{Technically speaking, there is a third possibility where, despite the existence of a semiclassical gravity dual, the naive holographic entanglement entropy formulae break down \cite{AP20}. This possibility is, however, not expected to be relevant in the setup considered in this paper. 
We would like to thank the JHEP referee for pointing this out. }

\begin{figure}[H]
    \centering
    \includegraphics[width=6cm]{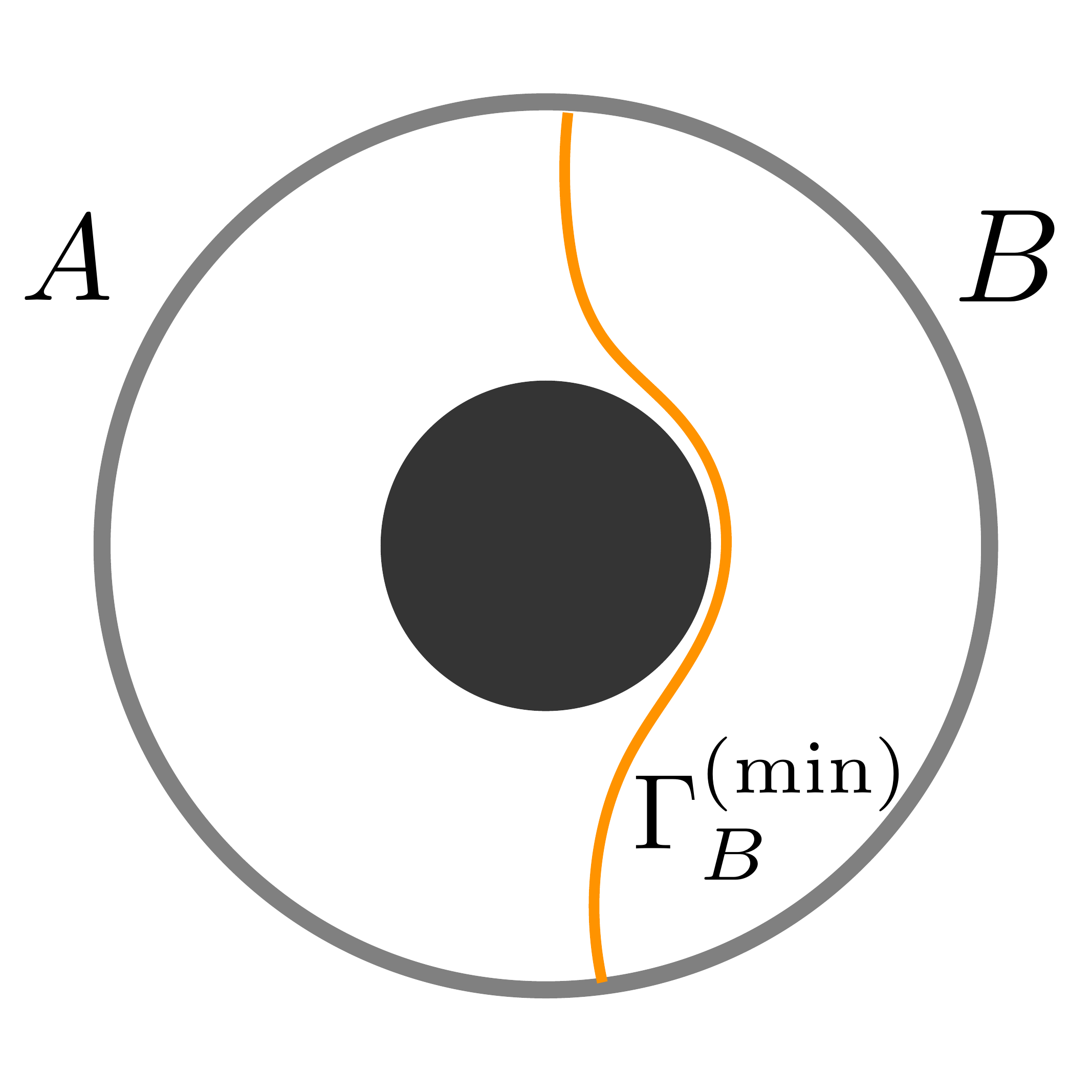}
    \hspace{1.5cm}
    \includegraphics[width=6cm]{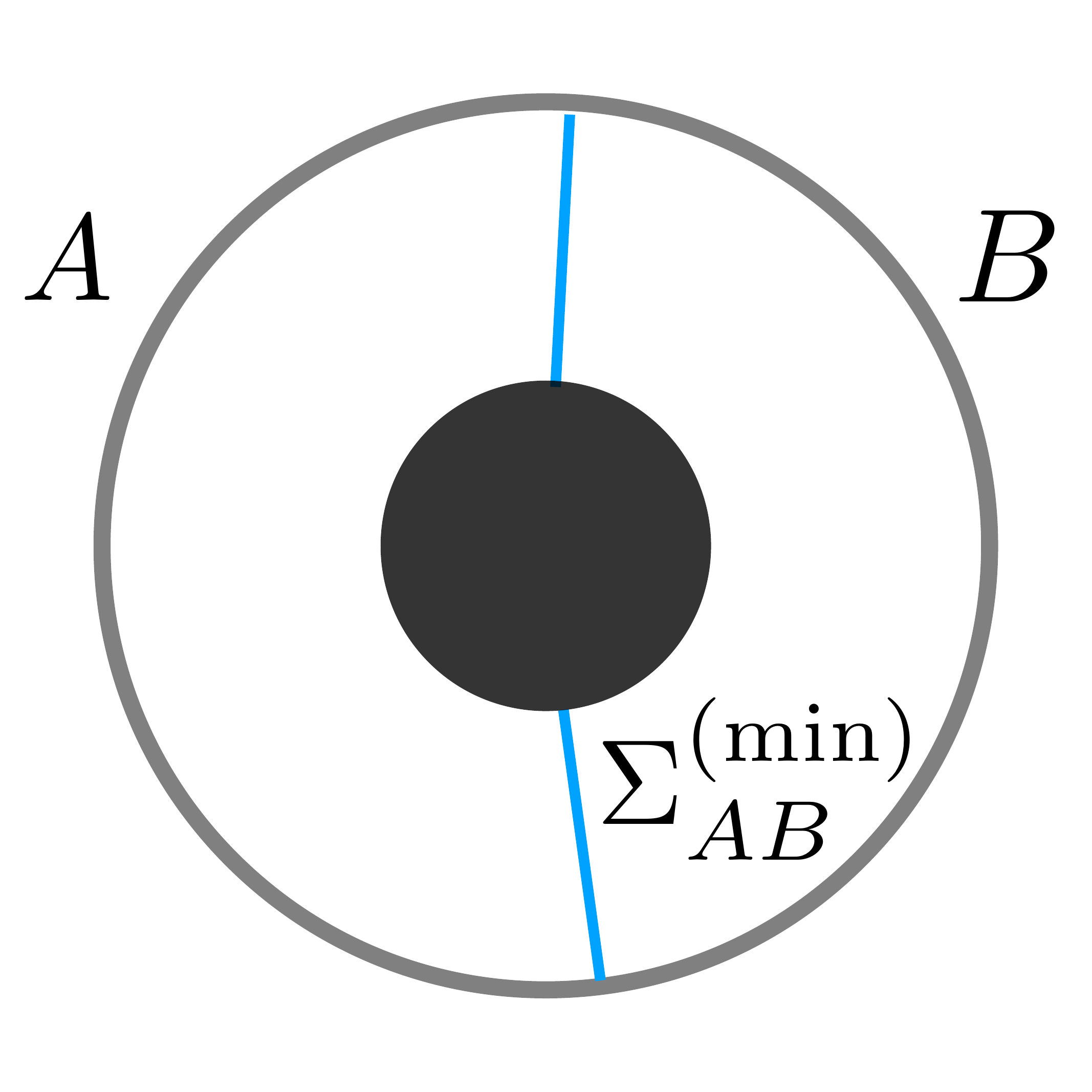}
    \caption{The geometry of an AdS black hole. The asymptotic boundary where the dual CFT lives is shown in gray. The spatial region of the CFT is divided into $A$ and $B$ where $B$ is slightly smaller than $A$. (Left) The orange surface $\Gamma_B^{\rm (min)}$ is the Ryu-Takayanagi surface that gives the entanglement entropy for a typical microstate and it wraps the black hole horizon. (Right) The blue surface $\Sigma_{AB}^{\rm (min)}$ gives $\frac{1}{2}S\left({\rm Tr}_{AA^*} |{\rm TFD}\rangle\langle{\rm TFD}| \right)$ and coincides the entanglement wedge cross section for a mixed thermal state.}
    \label{fig:BH}
\end{figure}

More precisely, for a typical AdS black hole state $|{\rm typ}\rangle$, when dividing the corresponding CFT into two regions $A$ and $B$ where $B$ is slightly smaller than $A$, the von Neumann entropy of the reduced density matrix on $B$ is the same as that of a canonical density matrix $\rho^{(\rm can)}$ with the corresponding temperature, i.e.
\begin{align}
    S\left({\rm Tr}_A|{\rm typ}\rangle\langle {\rm typ}|\right) \approx S\left({\rm Tr}_A \rho^{\rm (can)} \right)
\end{align}
at the leading order of $1/G_N$. Here, $G_N$ is the Newton constant in the AdS gravity, and the $G_N \rightarrow 0$ limit corresponds to the large degrees of freedom limit (large gauge rank number $N$ limit, large central charge $c$ limit, etc.) on the CFT side. Therefore, the von Neumann entropy of $\rho^{\rm (typ)}_B \equiv {{\rm Tr}_A |{\rm typ}\rangle\langle{\rm typ}|}$, i.e. the entanglement entropy between $A$ and $B$, can be evaluated by a Ryu-Takayanagi surface \cite{RT06,RT06b} or quantum extremal surface \cite{FLM13,EW15} wrapping the black hole horizon like the one shown in the left of figure \ref{fig:BH}. 

On the other hand, if a pure state has way much less entanglement between $A$ and $B$, then it will be atypical. Such an atypical state is called a disentangled state. To be more concrete for later discussions, let us define disentangled states in a rigorous way, following the one given in \cite{HP20}. 
\begin{definition*}[Disentangled States]
    A disentangled state $|{\rm disent}\rangle$ is a state which satisfies 
    \begin{align}\label{eq:disent_def}
       S\left({\rm Tr}_A|{\rm disent}\rangle\langle {\rm disent}|\right) \lesssim \frac{1}{2}S\left({\rm Tr}_{AA^*} |{\rm TFD}\rangle\langle{\rm TFD}| \right),
    \end{align}
    where ``$\lesssim$" means ``less than or equal to at the leading order of $1/G_N$". 
\end{definition*}
\noindent Here, $|{\rm TFD}\rangle$ is the thermofield double state 
\begin{align}
    |{\rm TFD}\rangle \propto \sum_i e^{-\beta E_i/2} |E_i\rangle |E^*_i\rangle,
\end{align}
where $|E_i\rangle$ is the energy eigenstate associated to the eigenvalue $E_i$, and $|E_i^*\rangle$ is its CPT conjugate living in a copy of the original Hilbert space. $A^* (B^*)$ is the copy of $A$ ($B$). Noting that $|{\rm TFD}\rangle$ is a symmetric purification of the canonical density matrix $\rho^{({\rm can})}$, 
\begin{align}
    \rho^{({\rm can})} = {\rm Tr}_{A^*B^*} |{\rm TFD}\rangle \langle{\rm TFD}|,
\end{align}
it directly follows from the positivity of the mutual information that 
\begin{align}
    S\left({\rm Tr}_A \rho^{\rm (can)} \right) - \frac{1}{2}S\left({\rm Tr}_{AA^*} |{\rm TFD}\rangle\langle{\rm TFD}|\right) 
    %\nonumber\\
    %= &\frac{1}{2} \left(S\left({\rm Tr}_{AA^*B^*} |\rm TFD\rangle\langle \rm TFD| \right) + S\left({\rm Tr}_{AA^*B} |\rm TFD\rangle\langle \rm TFD|\right) -S\left({\rm Tr}_{AA^*} |\rm TFD\rangle\langle \rm TFD| \right)   \right) \nonumber\\
    \geq  0. 
\end{align}
In a holographic CFT, the gap is $O(1/G_N)$ when the corresponding AdS black hole is sufficiently large. In this case, the entanglement held by a disentangled state is significantly smaller than a typical state, which is not sufficient to support an existence of a smooth horizon in its gravity dual \cite{HP20}. In a black hole geometry, the surface which gives $\frac{1}{2}S\left({\rm Tr}_{AA^*} |{\rm TFD}\rangle\langle{\rm TFD}|\right)$ ends on the black hole horizon as shown in the right of figure \ref{fig:BH}. 

Based on these, it is natural to ask how many disentangled states one can find in the sub-Hilbert space of the black hole microstates, which is spanned by the energy eigenstates in the energy window\footnote{Our arguments in this paper hold for any choice of the energy window as long as one can get a valid microcanonical ensemble from it. We will take $[E-\Delta E,E]$ just for convenience.} $[E-\Delta E,E]$. Let us denote this subspace as $\CH_{[E-\Delta E,E]}$ and call it the microcanonical subspace. 
In \cite{HP20}, Hayden and Penington showed that if extensive violations of additivity conjectures do not exist, then one can find $\exp\left[S_{\rm can}-O(1)\right]$ orthogonal disentangled states in a proper energy window $\CH_{[E-\Delta E,E]}$ at $G_N \rightarrow 0$, where $S_{\rm can}$ is the entropy computed from the corresponding canonical ensemble.\footnote{In \cite{HP20}, the Benkenstein-Hawking entropy $S_{BH}$ and the entropy computed from the corresponding canonical ensemble $S_{\rm can}$ are used interchangeably. In this paper, we would like to make a distinction between them for later discussions.} This is sufficient to account for the Bekenstein-Hawking entropy $S_{BH}$, since $S_{BH}$ matches the leading order of $S_{\rm can}$. Here, additivity conjectures are a series of equivalent conjectures in quantum information and communication theory \cite{Shor2004}. However, small violations of additivity conjectures have already been found \cite{Hastings09}, and there is so far no nontrivial analytic upper bound for possible violations of the additivity conjectures. In other words, their extensive violation may be even expectable. Possibilities of extensive violations are further discussed in, for example, \cite{JAVWY20,Wang22}. 
On the other hand, the existence of $\exp\left[S_{\rm can}-O(1)\right]$ orthogonal disentangled states in $\CH_{[E-\Delta E,E]}$ is surprising since it means that states without smooth horizons are sufficient to account for the black hole entropy at $G_N\rightarrow0$.
Therefore, one may argue that, since this is hard to believe, it may be possible to construct an explicit example of extensive violations of additivity conjectures from black holes. From this point of view, it is important to know how far one can go without any assumptions on additivity conjectures.
%{\blue In \cite{HP20}, Hayden and Penington showed that at least one of the following two statements must be true: (a) extensive violations of additivity conjectures \cite{Shor2004} exist; (b) one can find $\exp\left[S_{BH}\left(1-o(1)\right)\right]$ orthogonal disentangled states in $\CH_{[E-\Delta E,E]}$, which is sufficient to account for the Bekenstein-Hawking entropy $S_{BH}$ at $G_N \rightarrow 0$. Both of the statements are extraordinary. Here, additivity conjectures are a series of equivalent conjectures in quantum information and communication theory \cite{Shor2004}. However, small violations of them have already been found \cite{Hastings09}, and their extensive violation may be also expectable. Possibilities of extensive violations are further discussed in, for example, \cite{JAVWY20,Wang22}, but so far no such violation is found. Therefore, (a) would be a significant breakthrough in quantum information theory. On the other hand, (b) is surprising since it means that any black hole microstate can be written as a superposition of states without smooth horizons at $G_N\rightarrow0$. From this point of view, it is important to know how far one can go without any assumptions on additivity conjectures.} 

In this paper, we show that one can in fact find at least $\exp\left[S_{\rm micro} - f(S_{\rm micro})\right]$ disentangled states in $\CH_{[E-\Delta E,E]}$ , for any $f(x) \in \omega(1)$. Here, $\omega(1)$ means the set of functions which diverges faster than constants. For example, $\log x$ and $\log\log x$ are such functions. In other words, disentangled states are sufficient to account for the leading order of $S_{\rm micro}$ at $G_N\rightarrow0$ and the subleading correction can be arbitrarily small as long as it diverges. Accordingly, this can recover the Bekenstein-Hawking entropy $S_{BH}$ computed from the black hole area since it matches the leading order of $S_{\rm micro}$. Our proof has no assumptions on additivity conjectures. We start from the observation that the surface which gives $\frac{1}{2}S\left({\rm Tr}_{AA^*} |{\rm TFD}\rangle\langle{\rm TFD}|\right)$ (sketched in the right of figure \ref{fig:BH}) is exactly the same as a geometrical quantity called entanglement wedge cross section \cite{UT17, NDHZS17} in a black hole background, which can be related to a quantum informational quantity called entanglement of purification \cite{THLD02} in the corresponding holographic CFT.\footnote{It should be emphasized that the relation between entanglement wedge cross section and entanglement of purification \cite{UT17,NDHZS17}, on which our argument is based, is still a conjecture and has not yet been proven. The current stage of this conjecture will be reviewed in detail in section \ref{sec:EP=EW}.} This allows us to give an upper bound to the entanglement of formation \cite{BDSW96} of the microcanonical ensemble $\rho^{\rm (micro)}$ defined on $\CH_{[E-\Delta E,E]}$. This upper bound guarantees at least one disentangled state exists in $\CH_{[E-\Delta E,E]}$. Then we can exclude this disentangled state, consider its orthogonal complement and try to find another disentangled state in it. We will show that such a procedure can be repeated for at least $\exp\left[S_{\rm micro} - f(S_{\rm micro})\right]$ ($\forall f(x)\in\omega(1)$) times, which leads to at least $\exp\left[S_{\rm micro} - f(S_{\rm micro})\right]$ ($\forall f(x)\in\omega(1)$) times orthogonal disentangled states. The preliminaries for entanglement measures and holography used in the proof will be reviewed in section \ref{sec:preliminaries}, and the main statements will be presented as propositions and proven in section \ref{sec:hol_CFT}. 

In section \ref{sec:thermodynamic_limit}, we apply the techniques developed in section \ref{sec:hol_CFT} to an alternative situation. We consider an alternative type of atypical states, the area law states, at the standard thermodynamic limit\footnote{It is known that the $G_N \rightarrow 0$ limit in AdS/CFT shares many similar features with the standard thermodynamic limit (the large volume $V$ limit) in generic quantum many-body systems. For example, it is known that an $O(V^{1/2})$ correction \cite{MS19} appears at the phase transition point of the entanglement entropy found in generic quantum many-body systems satisfying the eigenstate thermalization hypothesis \cite{Srednicki94}, and one can find a similar $O(G_N^{-1/2})$ correction in AdS/CFT \cite{MWW20,DW20}. However, while being similar, they are two different limits and should be distinguished from each other. In a lattice regularization, one can intuitively regard $G_N\rightarrow0$ as the limit where the degrees of freedom on each site go to infinite, while the standard thermodynamic limit corresponds to the limit where the number of lattice itself goes to infinity.} $V\rightarrow \infty$, where $V$ is the system volume. Let us denote the system length as $L$ so that we have $V=O(L^d)$ where $d$ is the spatial dimension of the system. We will see that for any quantum many-body system, if the reflected entropy \cite{DF19} of its microcanonical ensemble follows an area law, i.e. scales as $O(L^{d-1})$, then one can find at least $\exp\left[S_{\rm micro}-O(\log S_{\rm micro})\right]$ orthogonal area law states in $\CH_{[E-\Delta E,E]}$. They can account for the thermodynamic entropy at the thermodynamic limit $L\rightarrow\infty$. Holographic CFTs\footnote{In holographic CFTs, the thermodynamic entropy coincides the Bekenstein-Hawking entropy $S_{BH}$ at the leading order. However, we would like to note that, when we talk about disentangled states, we are considering the $G_N \rightarrow 0$ limit with the volume $V$ fixed. On the other hand, when we talk about area law states, we are considering the $V \rightarrow \infty$ limit (with the Newton constant $G_N$ fixed if for holographic CFTs). While sharing similar features, these two limits, as well as the Landau symbols associated with them, should not be confused with each other.} are such examples whose reflected entropy satisfies the area law \cite{HM13,DF19}.
% We will also argue that generic quantum many-body systems with local interactions satisfy this assumption. 

In section \ref{sec:construction}, we present an explicit construction of $\exp\left[S_{\rm micro}-O(\log S_{\rm micro})\right]$ area law states. This is one of the crucial points in our paper: not only do we prove the existence, we even present an explicit construction! Moreover, we also discuss the possibility to apply the same method for constructing disentangled states.

In section \ref{sec:conclu}, we summarize our results, discuss their possible relations with additivity conjectures and atypical microstate counting, and show some future directions. 

{\bf Note added in arXiv v2:} Several improvements have been made compared to the arXiv v1 of this paper. Here, we list them out for readers' convenience.
\begin{enumerate}
    \item We have changed our notation for entropies to make a distinction between the Bekenstein-Hawking entropy $S_{BH}$ computed from the black hole geometry, the entropy computed from the canonical ensemble $S_{\rm can}$ and the entropy computed from the microcanonical ensemble $S_{\rm micro}$. This was not necessary in arXiv v1 since we only discussed the leading order for which they match each other. However, this turns out to be necessary in the current version in order to make the following improvements. 
    \item We have improved the subleading correction of our main results presented in Corollary \ref{coro:hol_final} and \ref{coro:qmb}. These are phrased in terms of $S_{\rm micro}$, and the related discussions have been modified throughout the paper.
    \item Based on the above improvement, appendix \ref{app:comparison} has been added to compare with a statement presented in \cite{HP20}.
    \item We have added a prerequisite to theorem \ref{thm:Increment_bound}, which was missed in arXiv v1 though it does not affect the validity of the following discussions. 
    The proof has also been improved. 
\end{enumerate}

\section{Preliminaries}\label{sec:preliminaries}
As we have introduced in the previous section, upper bounding the entanglement of formation of microcanonical ensembles in holographic CFTs will be an important piece of our jigsaw puzzle. In this section, we review several preliminaries for this purpose. We will firstly review entanglement of formation and other relevant correlation measures in general quantum systems in section \ref{sec:entanglement_measures}. Then we will review entanglement of purification in holography and gravity duals of microcanonical ensembles in section \ref{sec:EP=EW} and \ref{sec:hol_micro} respectively. Readers who are familiar with these topics can skip this section.

\subsection{Entanglement of formation and other correlation measures}\label{sec:entanglement_measures}

Consider a bipartite system composited by $A$ and $B$, and denote the Hilbert space associated with it as $\CH_A \otimes \CH_B$. For a (possibly mixed) state $\rho \in \CD(\CH_A\otimes\CH_B)$, there are many quantities to measure the entanglement and correlation between $A$ and $B$. 

Since entanglement is an important resource which classical systems does not have, one may want to construct a function which counts only the amount of entanglement but not any classical correlations between $A$ and $B$. Such a function is called an entanglement measure. More precisely, an entanglement measure must satisfy some axioms \cite{DHR02,PV07,HHH09,BCDS18}, which guarantees that it can quantitatively measure the amount of entanglement and does not contain any classical correlations,
which can be generated via LOCC operations.
See appendix \ref{app:entanglement_measure} for a set of axioms required for an entanglement measure. On the other hand, functions which do not fully satisfy all the axioms are often called correlation measures. In the following, let us list some entanglement measures and correlation measures which will be used in this paper. 

\paragraph{Entanglement entropy}~\par
When $\rho$ is a pure state, the entanglement entropy is defined as 
\begin{align}
    E_{vN}(\rho) = S(\rho_A) = S(\rho_B),
\end{align}
where 
\begin{align}
    \rho_{A(B)} \equiv {\rm Tr}_{B(A)} (\rho),
\end{align}
is the reduced density matrix on $\CH_{A(B)}$ and 
\begin{align}
    S(\sigma) = -{\rm Tr} \left(\sigma \log{\sigma}\right)
\end{align}
% is the von Neumann entropy, is the only function which satisfy all the axioms for entanglement measures \cite{DHR02}. Therefore, entanglement entropy is an entanglement measure when restricted to pure states.
is the von Neumann entropy. The reduced von Neumann entropy on pure states is the only function which satisfy all the axioms for entanglement measures \cite{DHR02}. Therefore, the entanglement entropy is an entanglement measure when restricted to pure states.

On the other hand, constructing a computable entanglement measure for a mixed state which has a good operational meaning is much more challenging. 

\paragraph{Entanglement of formation}~\par
One way to define a correlation measure for a mixed state $\rho$ is to regard $\rho$ as an ensemble of pure states 
\begin{align}
    \rho = \sum_i p_i |\psi_i \rangle \langle \psi_i|, 
\end{align}
and then consider the average entanglement entropy over this ensemble:
\begin{align}
    \sum_i p_i E_{vN}(|\psi_i \rangle \langle \psi_i|). 
\end{align}
The entanglement of formation \cite{BDSW96} of $\rho$ is defined as the minimal one of such average entanglement entropies over all variations of possible ensemble decomposition:
\begin{align}\label{eq:eof_def}
    E_F(\rho) \equiv \inf \left\{ \sum_i p_i E_{vN}(|\psi_i \rangle \langle \psi_i|) : \rho = \sum_i p_i |\psi_i \rangle \langle \psi_i| \right\}.
\end{align}
The entanglement of formation is not an entanglement measure, and it is usually not computable since the definition includes a minimization. However, it has a good operational meaning related to quantum measurements \cite{BDSW96} which motivates people to consider it. 

In this paper, the scaling behavior of $E_F$ in holographic CFTs and quantum many-body systems will play a central role. Therefore, we will review two correlation measures which give rigorous upper bounds to $E_F(\rho)$. 

\paragraph{Entanglement of purification}~\par 
Another idea to measure the correlation between $A$ and $B$ for a mixed state $\rho$ is to first uplift it into a pure state and then consider entanglement entropy on that pure state. Such an uplifting procedure is called purification. More precisely, for an arbitrary ancilla $C$ and a pure state $|\psi_{ABC}\rangle$ in $\CH_A \otimes \CH_B \otimes \CH_C$, if it reduces to $\rho \in \CD(\CH_A \otimes \CH_B)$ after tracing out $C$, i.e. 
\begin{align}
    \rho = {\rm Tr}_C |\psi_{ABC}\rangle \langle \psi_{ABC}|, 
\end{align}
then $|\psi_{ABC}\rangle$ is called a purification of $\rho$. By artificially dividing $\CH_C$ into two parts
\begin{align}
    \CH_C = \CH_{A'} \otimes \CH_{B'},
\end{align}
and assigning $A'(B')$ to $A(B)$, we can get $A\cup A'~(B\cup B')$ as an extension of $A(B)$. It is natural to expect that the entanglement entropy between $A\cup A'$ and $B\cup B'$, i.e. $S\left({\rm Tr}_{BB'}|\psi_{ABC}\rangle \langle \psi_{ABC}|\right)$ can measure the correlation between $A$ and $B$. 

Obviously, different choices of $|\psi_{ABC}\rangle$ and different ways of factorizing $\CH_{C}$ give different results. Among all these choices, the minimal entanglement entropy evaluated in this way is called the entanglement of purification \cite{THLD02}, 
\begin{align}\label{eq:eop_def}
    E_P(\rho) \equiv \inf \Bigg\{ S\left({\rm Tr}_{BB'}|\psi_{ABC}\rangle \langle \psi_{ABC}|\right) : \rho = {\rm Tr}_C |\psi_{ABC}\rangle \langle \psi_{ABC}| ~ {\rm and} ~ C=A'\cup B' \Bigg\}.
\end{align}
$E_P(\rho)$ gives a rigorous upper bound of entanglement of formation $E_F(\rho)$ as \cite{THLD02}
\begin{align}\label{eq:eof<eop}
    E_F(\rho) \leq E_P(\rho).
\end{align}
On the other hand, while $E_P(\rho)$ is not computable in general due to the minimization, it can be evaluated in a class of states in holographic CFTs \cite{UT17,NDHZS17}. We will apply these two properties when discussing black hole microstates later. 

\paragraph{Reflected entropy}~\par
Since the entanglement of purification is not computable in general, one may want to consider a specific way of purification with which one can get a computable quantity. For example, let us consider a mixed state $\rho \in \CD(\CH_A \otimes \CH_B)$ and diagonalize it as 
\begin{align}
    \rho = \sum_\nu p_\nu |\varphi_\nu\rangle \langle \varphi_\nu|.
\end{align}
For each $|\varphi_\nu\rangle \in \CH_A \otimes \CH_B$, we can consider its CPT conjugate and denote it as $|\varphi_\nu^*\rangle \in \CH_{A^*} \otimes \CH_{B^*}$. Then, obviously, the state 
\begin{align}
    |\sqrt{\rho}\rangle = \sum_\nu \sqrt{p_\nu} |\varphi_\nu\rangle |\varphi_\nu^*\rangle \label{eq:can_purification}
\end{align}
defined on $\CH_A \otimes \CH_B \otimes \CH_{A^*} \otimes \CH_{B^*}$ serves as a purification of $\rho$. Such a purification is called the canonical purification \cite{DF19}. 

Associated with the canonical purification, the reflected entropy \cite{DF19} is defined as the entanglement entropy between $A\cup A^*$ and $B\cup B^*$:
\begin{align}
    E_{RE}(\rho) = S({\rm Tr}_{BB^*}|\sqrt{\rho}\rangle \langle \sqrt{\rho} | ). 
\end{align}
By definition, it satisfies 
\begin{align}
    E_{P}(\rho) \leq E_{RE}(\rho). 
\end{align}
Therefore, combining with \eqref{eq:eof<eop}, we have 
\begin{align}\label{eq:F_P_RE}
    E_{F}(\rho) \leq E_{P}(\rho) \leq E_{RE}(\rho). 
\end{align}
The best advantage of $E_{RE}(\rho)$ is that it is usually easy to compute. We will apply these features to discuss generic quantum many-body systems in the thermodynamic limit. 

\subsection{Entanglement of purification in AdS/CFT}\label{sec:EP=EW}
While the entanglement of purification is in general very hard to compute, it can be easily evaluated for geometric states in holographic CFTs. Here, a holographic CFT is a CFT which admits a semiclassical gravity dual via the AdS/CFT correspondence \cite{Maldacena97}. However, not all the states in $\CH_{\rm CFT}$ necessarily admit a geometrical dual\footnote{For example, it is believed a superposition of a geometric state corresponding to the vacuum AdS and another state corresponding to an AdS black hole does not admit a geometrical dual.}. A state which admits a geometrical dual is called a geometric state. 

\begin{figure}[H]
    \centering
    \includegraphics[width=6cm]{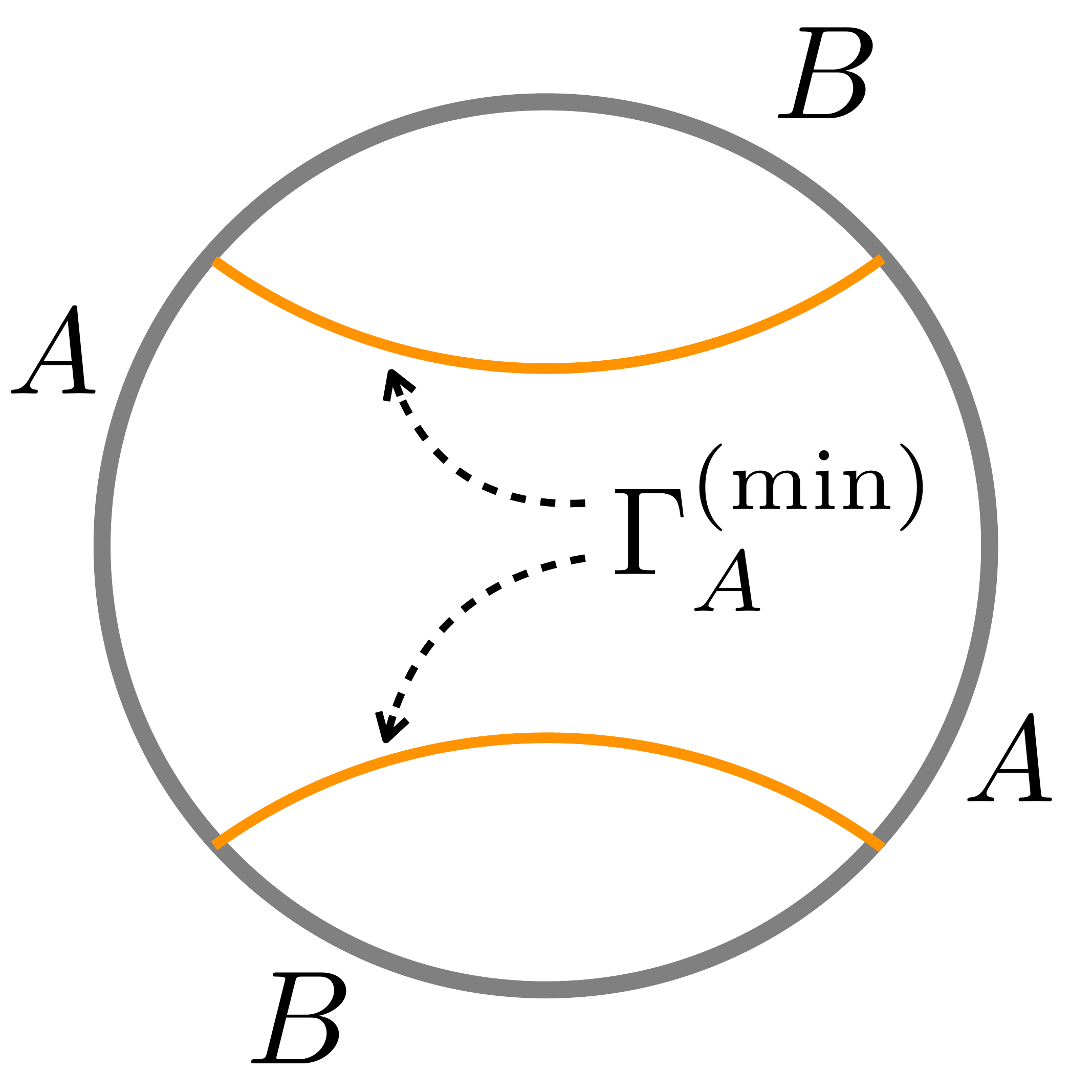}
    \hspace{1.5cm}
    \includegraphics[width=6cm]{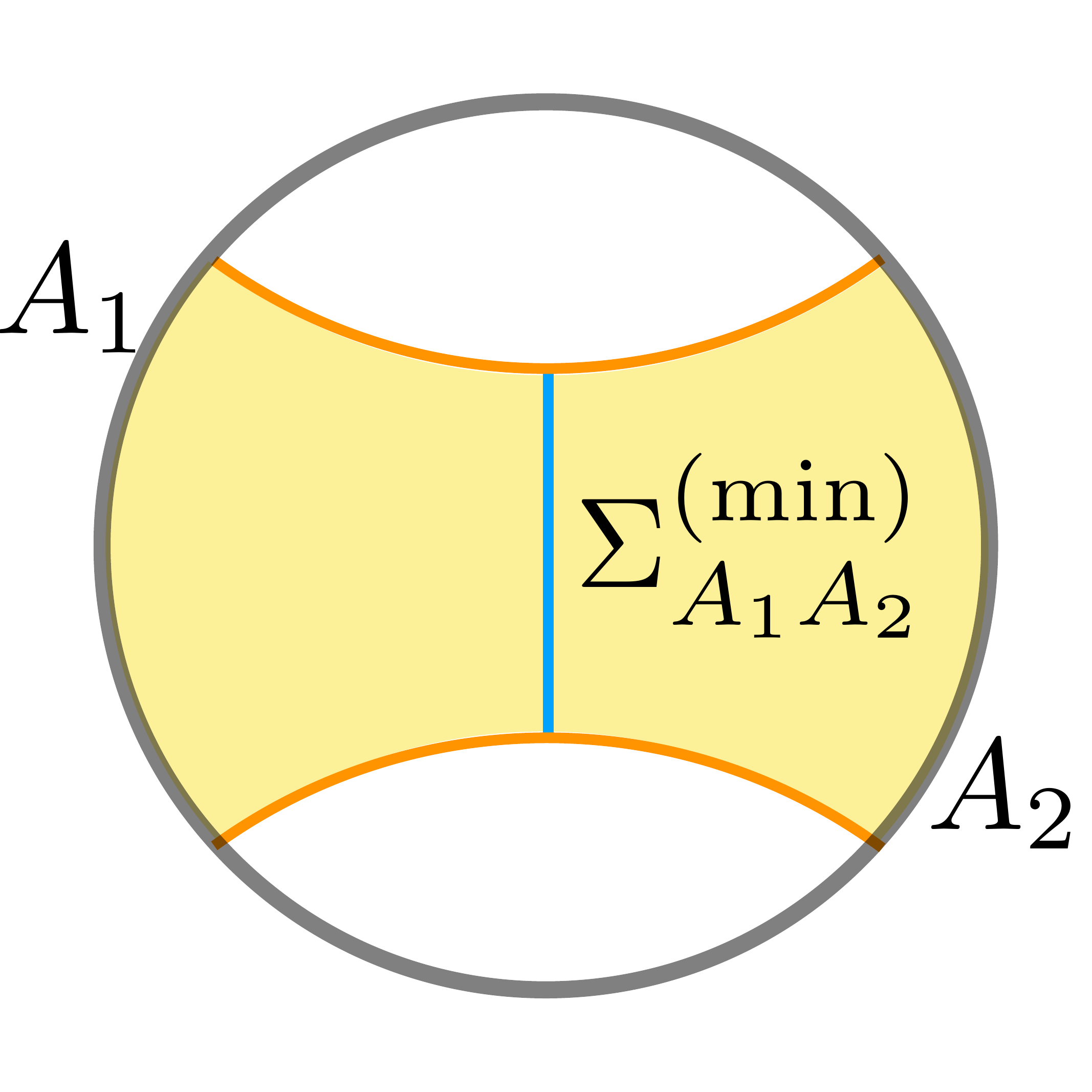}
    \caption{A time slice of a global AdS$_3$. The asymptotic boundary where the dual CFT lives is shown in gray. The spatial region of the CFT is divided into $A$ and $B$ where $A$ is the two segments on the left and right, and $B$ is the two segments on the top and the bottom. (Left) The orange surface $\Gamma_A^{\rm (min)}$ shows the Ryu-Takayanagi surface. (Right) The entanglement wedge of $A=A_1\cup A_2$ is shaded in yellow. The surface $\Sigma_{A_1A_2}^{\rm (min)}$ which gives the entanglement wedge cross section $E_W(\rho_A, A_1 : A_2)$ is shown in blue.}
    \label{fig:RT}
\end{figure}

For geometric states, the entanglement entropy and the entanglement of purification correspond to geometric objects in the AdS gravity. Consider a static geometric state $\rho$ in the AdS$_{d+2}$/CFT$_{d+1}$ correspondence and divide the whole $d$-dimensional spatial region of the CFT into $A$ and $B$. The von Neumann entropy of the reduced density matrix on $A$ is given by the Ryu-Takayanagi formula \cite{RT06,RT06b}
\begin{align}\label{eq:RT}
    S\left({\rm Tr}_B (\rho) \right) = \min_{\Gamma_A} \frac{{\rm Area}(\Gamma_A)}{4G_N},
\end{align}
where $\Gamma_A$ is a $d$-dimensional surface which is homologous to $A$. The $\Gamma_A$ with the minimal area is called the RT surface. See the left of figure \ref{fig:RT} for a sketch. 

Here we would like to note that, since we are considering the CFT, the Hilbert space factorization and the entanglement entropy are not well-defined in a strict sense. The formula above should be understood in the following way. Assuming there exists a lattice regularization of the CFT under consideration, one can define and compute the entanglement entropy. The entanglement entropy computed in this way will depend on the UV cutoff $\epsilon$ (or the lattice distance) and get divergent when taking $\epsilon$ to zero. Throughout this paper, we will assume that there exists a lattice regularization, and we will discuss the scaling behavior of various quantities such as the entanglement entropy against large parameters such as $1/G_N$, with the UV cutoff $\epsilon$ being fixed.

The region surrounded by $A$ and $\Gamma_A^{\rm (min)}$ is called the entanglement wedge\footnote{More precisely, what we are considering here is the time-reflection symmetric time slice of the entanglement wedge.} \cite{CKNvR12,Wall12,HHLR14} of $A$ and it is considered to be the gravity dual of the reduced density matrix 
\begin{align}
    \rho_{A} \equiv {\rm Tr}_B (\rho). 
\end{align}
If we divide $A$ into $A = A_1 \cup A_2$, we can consider a quantity called the entanglement wedge cross section\footnote{In the following, we will sometimes explicitly show the bipartition if there might be confusion.}
\begin{align}
    E_W(\rho_A, A_1 : A_2) \equiv \min_{\Sigma_{A_1 A_2}} \frac{{\rm Area}(\Sigma_{A_1 A_2})}{4G_N}
\end{align}
where $\Sigma_{A_1 A_2}$ is a $d$-dimensional surface which divide the entanglement wedge into two parts with $A_1$ and $A_2$ living in each part.

It is conjectured in \cite{UT17,NDHZS17} that the entanglement of purification between $A_1$ and $A_2$ of state $\rho_A$ equals the entanglement wedge cross section $E_W(\rho_A, A_1 : A_2)$:
\begin{align}\label{eq:hol_EoP}
    E_P(\rho_A, A_1 : A_2) = E_W(\rho_A, A_1 : A_2).
\end{align}
This conjecture is justified in \cite{UT17,NDHZS17} by a two-step argument. According to the surface/state correspondence \cite{MT15} or the tensor network description of holography \cite{Swingle09,PYHP15,HNQTY16}, the entanglement wedge with some ancilla degrees of freedom put on the RT surface can be regarded as an ``optimal geometric purification", and hence finding the minimal $\Gamma_{A_1A_2}$ dividing the RT surface $\Gamma_{A}$ into two parts corresponds to finding the optimal way to divide the ancilla in this purification. This first step suggests that $E_W$ at least gives an upper bound of $E_P$, i.e. $E_P \leq E_W$. As the second step, one can geometrically prove $E_W$ satisfy all the known inequalities satisfied by $E_P$ \cite{UT17,NDHZS17}. 

This conjecture between the entanglement of purification and the entanglement wedge cross section plays a central role in our discussion about black hole microstates. In fact, we only need a weaker assumption $E_P \leq E_W$. For later reference, let us summarize this as the following statement. 

\begin{customconj}{0}\label{conjecture0}
    Consider a holographic CFT,
    and divide the spatial region into $A = A_1 \cup A_2$ and its complement.
    For a geometric state which admits a semiclassical gravity dual, the entanglement of purification $E_P(\rho_A,A_1:A_2)$ is upperbounded by the entanglement wedge cross section $E_W(\rho_A, A_1:A_2)$, i.e.
    \begin{align}
        E_P(\rho_A,A_1:A_2) \leq E_W(\rho_A, A_1:A_2).
    \end{align}
\end{customconj}

\subsection{Holographic dual of the microcanonical ensemble}\label{sec:hol_micro}
As we have explained above, not all the states in a holographic CFT are geometric states. Since the standard AdS/CFT dictionary \cite{GKP98,Witten98} is an equivalence between the partition functions of the gravity path integral and the CFT path integral, when a state can be easily prepared via a path integral, then one can use the semiclassical approximation of the gravity path integral, by picking up the configuration which minimizes the gravitational action $I_{\rm grav}$. 

Geometric states constructed in this way include the ground state, canonical ensembles 
\begin{align}
    \rho^{\rm (can)}(\beta) \propto \sum_i e^{-\beta E_i} |E_i\rangle\langle E_i|,
\end{align}
thermofield double states (i.e. canonical purifications of canonical ensembles)
\begin{align}\label{eq:TFD}
    |{\rm TFD}(\beta)\rangle \propto \sum_i e^{-\beta E_i/2}|E_i\rangle|E_i^*\rangle,
\end{align}
and so on. Here, $E_i$ is the $i$-th eigenvalue of the Hamiltonian $H$ and $|E_i\rangle$ is the corresponding eigenstate. These three examples correspond to global AdS, one-sided AdS black holes, and two sided AdS black holes \cite{Maldacena01} respectively. 

In this paper, we would like to focus on microcanonical ensembles
\begin{align}\label{eq:micro_ensemble}
    \rho^{\rm (micro)}(E) \propto \sum_{i {\rm ~s.t.}~E_i \in [E-\Delta E,E]} |E_i\rangle\langle E_i|, 
\end{align}
which are more complicated to prepare with a standard path integral. Fortunately, the gravity dual of the canonical purifications of \eqref{eq:micro_ensemble}, named microcanonical thermofield double states, 
\begin{align}\label{eq:MCTFD}
    |{\rm MCTFD}(E)\rangle \propto \sum_{i {\rm ~s.t.}~E_i \in [E-\Delta E,E]} |E_i\rangle |E_i^*\rangle
\end{align}
has been already discussed in \cite{Marolf18}\footnote{In fact, the exact setup analyzed in \cite{Marolf18} is $|\psi\rangle \propto \sum_{E'} e^{-\beta E'} f(E'-E) |E'\rangle |E'^*\rangle$ where $f(E'-E) = \exp((E'-E)^2/4\sigma^2)$. However, as commented in \cite{Marolf18}, the results hold for any $f(E'-E)$ concentrated around $E'=E$.}. This is again given by a two-sided black hole in the semiclassical limit $G_N\rightarrow0$. Accordingly, the gravity dual of a microcanonical ensemble is given by a one-sided black hole. 

The difference between the gravity duals of a canonical ensemble and a microcanonical ensemble is that, while we need to find the classical geometry which minimize the gravitational action $I_{\rm grav}$ on the AdS side when using a standard path integral to prepare a canonical ensemble or a thermofield double state, we need to find the one which maximize another functional, which coincides the black hole entropy for black hole geometries, when considering the microcanonical counterparts \cite{Marolf18}.

The results explained above can be naturally understood from the view point of the relation between statistical mechanics and thermodynamics. In statistical mechanics, different ensembles including the canonical ensemble and the microcanonical ensemble are equivalent to each other in reproducing the macroscopic quantities at the limit of large degrees of freedom (i.e. the thermodynamic limit). In the current case, the semiclassical limit $1/G_N\rightarrow\infty$, where $G_N$ is the Newton constant, plays a similar role of the thermodynamic limit. At the thermodynamic limit, the natural thermodynamic potential associated with the canonical ensemble is the Helmholtz free energy, which holds temperature $1/\beta$ as a natural variable\footnote{Here, ``a thermodynamic potential $A$ holds $x,y,\cdots$ as natural variables" means that all the thermodynamic quantities can be determined from $A$ as a function of $x,y,\cdots$. For example, when expressing the Helmholtz free energy $F$ as a function of the temperature $1/\beta$ and the volume $V$, all the thermodynamic quantities can be determined from partial derivatives of $F[1/\beta,V]$. However, when expressing $F$ as a function of the energy $E$ and the volume $V$, some information is lost in $F(E, V)$. A thermodynamic function expressed by its natural variables is called a complete thermodynamic function in some textbooks. Complete thermodynamic functions are equivalent to each other under Legendre transformations.}. On the other hand, the natural thermodynamic potential associated with the microcanonical ensemble is the thermodynamic entropy, which holds energy $E$ as the natural variable. In the current case, the gravitational action $I_{\rm grav}$ parameterized by $\beta$ corresponds to the Helmholtz free energy, and the functional which coincides the black hole entropy in black hole geometries \cite{Marolf18} parameterized by $E$ corresponds to the thermodynamic entropy. The thermodynamic equilibrium is realized when the Helmholtz free energy is minimized (or equivalently, when the thermodynamic entropy is maximized). 

Moreover, it is also shown in \cite{Marolf18} that the Ryu-Takayanagi formula \cite{RT06,RT06b} and its covariant extension, the Hubeny-Rangamani-Takayanagi formula \cite{HRT07} still hold in the microcanonical counterparts. Therefore, it is plausible that we can also apply the $E_P = E_W$ conjecture (or at least $E_P \leq E_W$) for microcanonical ensembles. 

\section{Disentangled black hole microstates in AdS/CFT}\label{sec:hol_CFT}

In this section, we show that, for a holographic CFT, there exists an orthogonal basis of the microcanonical subspace $\CH_{[E-\Delta E, E]}$ such that, for any $f(x)\in \omega(1)$, at least $\exp\left[S_{\rm micro}-f(S_{\rm micro})\right]$ elements are disentangled states. Here $S_{\rm micro} \equiv \log {\rm dim} \CH_{[E-\Delta E, E]}$ is the entropy of the corresponding AdS black hole computed from the microcanonical ensemble. Since $S_{\rm micro}$ and the Bekenstein-Hawking entropy $S_{BH}$ match at the leading order, we can say that counting orthogonal disentangled states can give $S_{BH}$ at the leading order.

We will firstly give an upper bound to the entanglement of formation of the microcanonical ensemble $\rho^{\rm (micro)}$ using holography in section \ref{sec:EoF_bound}. This bound indicates that there is at least one disentangled state in $\CH_{[E-\Delta E, E]}$. Then we can exclude this state and consider its orthogonal complement (and a new microcanonical density matrix associated to it). It turns out the entanglement of formation of the new microcanonical ensemble can also be upper bounded. We will show some results which are crucial in this argument in section \ref{sec:general_results} merely from linear algebra. This means results in section \ref{sec:general_results} are universal for any quantum systems. According as the new upper bound behaves, we may repeat the above procedure again and again. We will finally show this procedure can be repeated $\exp\left[S_{\rm micro}-f(S_{\rm micro})\right]$ times for any $f(x)\in \omega(1)$ in section \ref{sec:hol_final_statement}. 

\subsection{Upper bounding \texorpdfstring{$E_F$}{EoF} from holography}\label{sec:EoF_bound}
As the first step, we will use the preliminaries reviewed in section \ref{sec:preliminaries} to give an upper bound to $E_F$ of $\rho^{\rm (micro)}$. We consider dividing the spatial region of the CFT into two parts $A$ and $B$, where $B$ is slightly smaller than $A$. See figure \ref{fig:BH} for a sketch. 

\begin{figure}
    \centering
    \includegraphics[width=6cm]{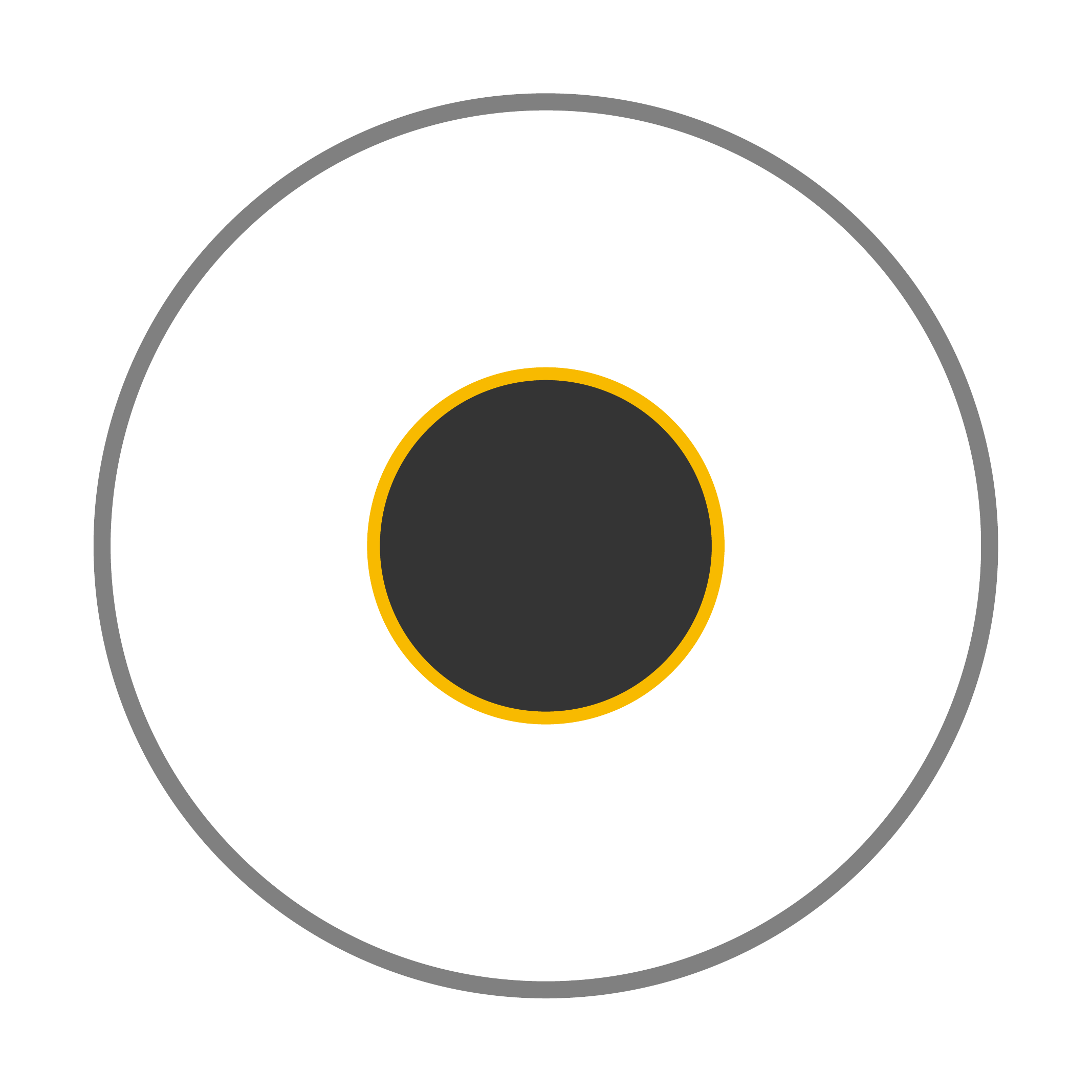}
    \hspace{1.5cm}
    \includegraphics[width=6cm]{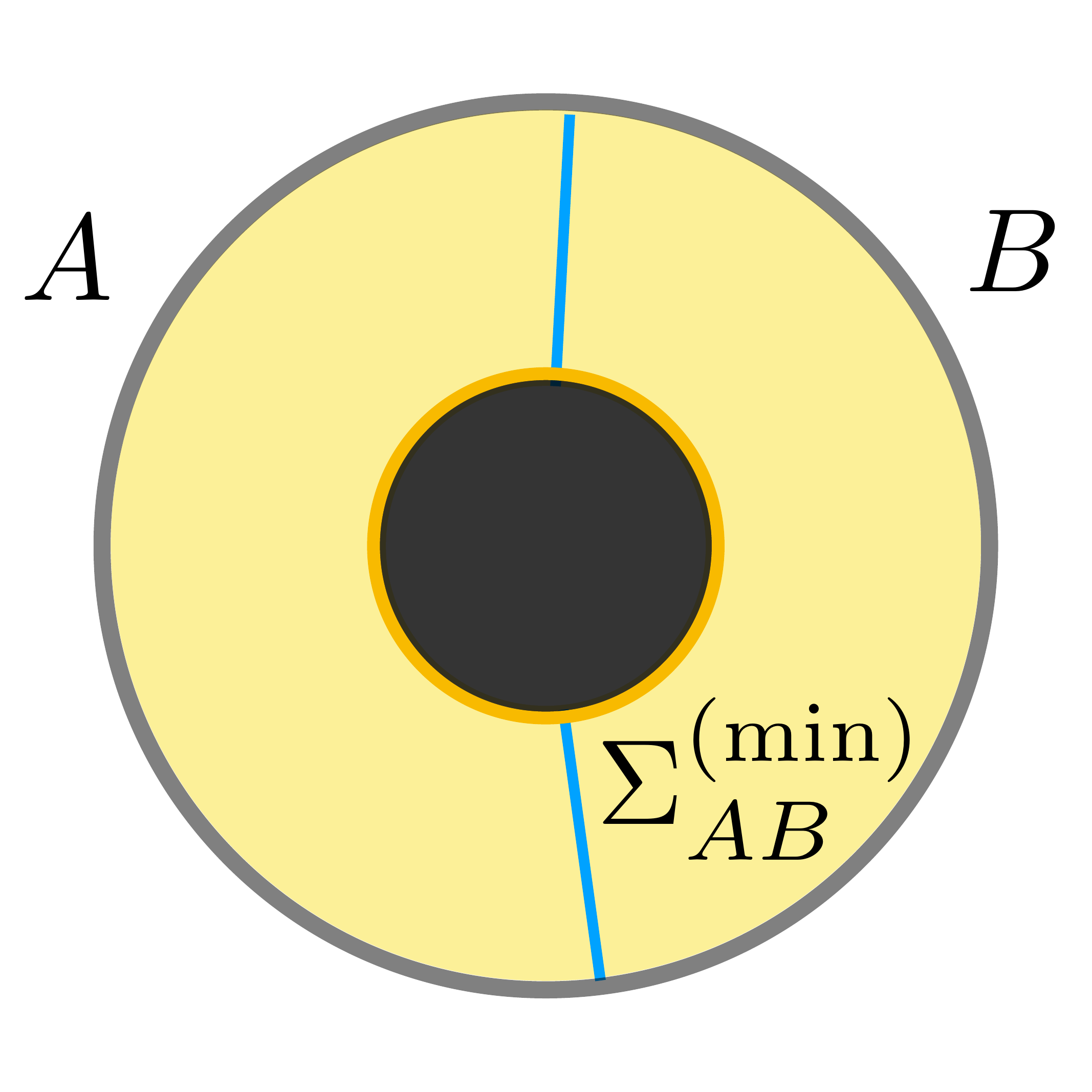}
    \caption{AdS black hole as the gravity dual of the microcanonical density matrix $\rho^{\rm (micro)}$.
    (Left) The orange surface, which coincides the black hole horizon shows the Ryu-Takayanagi surface of the whole system.
    (Right) The entanglement wedge of the whole system is the outside region of the black hole shown in yellow. Dividing the spatial region of the CFT into $A$ and $B$, the blue surface $\Sigma_{AB}^{\rm (min)}$ gives the entanglement wedge cross section $E_W(\rho^{\rm (micro)}, A : B)$.}
    \label{fig:micro}
\end{figure}

As we have reviewed in section \ref{sec:hol_micro}, the gravity dual of $\rho^{\rm (micro)}$ is a black hole geometry. The entanglement wedge of the whole system $A\cup B$ is the exterior of the black hole, as shown in figure \ref{fig:micro}. Accordingly, the entanglement wedge cross section $E_W$ is given by the area of the minimal surface connecting the interface $\partial A = \partial B$ and the black hole horizon. Let us denote this minimal surface as $\Sigma^{\rm (min)}_{AB}$. Combining the holographic formula for entanglement of purification \eqref{eq:hol_EoP} and the general relation between entanglement of formation and entanglement of purification \eqref{eq:eof<eop}, we have 
\begin{align}
    E_F \leq E_P \leq E_W = \frac{{\rm Area}(\Sigma^{\rm (min)}_{AB})}{4G_N} + O(1).
\end{align}
Therefore, by definition of $E_F$, it is straightforward to say that there exists an ensemble decomposition of $\rho^{\rm (micro)}$ such that the averaged entanglement entropy is less than or equal to ${\rm Area}(\Sigma^{\rm (min)}_{AB})/4G_N$ at leading order. Let us summarize this as a proposition for later reference. 

%\newpage
\begin{customprop}{1}\label{prop:hol_bound}
    In a holographic CFT, let $\rho^{\rm (micro)}$ be a microcanonical ensemble and divide the spatial region into $A$ and $B$ where $B$ is slightly smaller than $A$. Then there exists a ensemble of states $\{(p_i,|\psi_i\rangle)\}_{i=1,2,\cdots}$ such that 
    \begin{align}
        \rho^{\rm (micro)} = \sum_i p_i |\psi_i\rangle \langle\psi_i|, 
    \end{align}
    and the averaged entanglement entropy 
    \begin{align}\label{eq:EvN_ave_hol}
        \overline{E_{vN}}(\{(p_i,|\psi_i\rangle)\}_{i=1,2,\cdots}) = \sum_{i} p_i E_{vN}(|\psi_i \rangle \langle \psi_i|) \lesssim \frac{{\rm Area}(\Sigma^{\rm (min)}_{AB})}{4G_N},
    \end{align}
    where ``$\lesssim$" means ``less than or equal to at leading order".
\end{customprop}

Noting that ${\rm Area}(\Sigma^{\rm (min)}_{AB})/4G_N$ and  $\frac{1}{2}S\left({\rm Tr}_{AA^*} |{\rm TFD}\rangle\langle{\rm TFD}| \right)$ matches at the leading order of $1/G_N$, according to the definition of disentangled states \eqref{eq:disent_def}, it straightforwardly follows from the positivity of probability and von Neumann entropy that at least one element $|\psi_j\rangle$ with $p_j \neq 0$ in $\{|\psi_i\rangle)\}_{i=1,2,\cdots}$ is a disentangled state, i.e. it satisfies
\begin{align}
        E_{vN}(|\psi_j \rangle \langle \psi_j|) \lesssim \frac{{\rm Area}(\Sigma^{\rm (min)}_{AB})}{4G_N} \approx \frac{1}{2}S\left({\rm Tr}_{AA^*} |{\rm TFD}\rangle\langle{\rm TFD}|\right). 
\end{align}
However, to this point, it is nontrivial whether this $|\psi_j\rangle$ is in the microcanonical subspace or not. It will soon be shown in proposition \ref{lemma} that this $|\psi_j\rangle$ is indeed in the microcanonical subspace.

\subsection{General results for microcanonical ensembles}\label{sec:general_results}
In this subsection, we show some general properties of microcanonical density matrices from linear algebra, and comment on their applications on the system we are considering. Note that the results presented in this subsection are general, and no properties of specific physical systems will be used to show them.

\begin{customprop}{2}\label{lemma}
Let $\rho$ be a density matrix which can be written as
$\rho = \Pi_{\mathcal{E}}/W$,
where $\Pi_{\mathcal{E}}$ is the projection operator
onto a $W$-dimensional subspace $\mathcal{E} \cong \mathbb{C}^{W}$. 
If an ensemble of states $\{(p_i,|\psi_i\rangle)\}_{i=1,2,\cdots}$
generates $\rho$, i.e.,
\begin{align}
  \rho = \sum_i p_i |\psi_i\rangle\langle{\psi_i}|, \label{eq:DM_as_ensemble}
\end{align}
then for every $|\psi_j\rangle \in \{|\psi_i\rangle\}_{i=1,2,\cdots}$, $|\psi_j\rangle \in \mathcal{E}$ if $p_j \neq 0$.
\end{customprop}

\begin{proof}
Note that
\begin{align}
  p_j \|(1-\Pi_{\mathcal{E}})|\psi_j\rangle\|^4
  &= p_j |\langle\psi_j|(1-\Pi_{\mathcal{E}})|\psi_j\rangle|^2
  \leq \sum_i p_i |\langle\psi_i|(1-\Pi_{\mathcal{E}})|\psi_j\rangle|^2.
\end{align}
Therefore, using \eqref{eq:DM_as_ensemble}, we have
\begin{align}
  p_j \|(1-\Pi_{\mathcal{E}})|\psi_j\rangle\|^4
  &\leq \langle\psi_j|(1-\Pi_{\mathcal{E}})\rho(1-\Pi_{\mathcal{E}})|\psi_j\rangle
  = \langle\psi_j|(1-\Pi_{\mathcal{E}})\frac{\Pi_{\mathcal{E}}}{W}(1-\Pi_{\mathcal{E}})|\psi_j\rangle
  = 0.
\end{align}
This implies that $|\psi_j\rangle = \Pi_{\mathcal{E}} |\psi_j\rangle \in \mathcal{E}$ or $p_j = 0$.
\end{proof}

Based on proposition \ref{lemma}, we can say at least one $|\psi_{j}\rangle$ appearing in proposition \ref{prop:hol_bound} is a disentangled state in $\CH_{[E-\Delta E, E]}$.
Therefore, we can consider
% excluding the orthogonal complement of $\psi_{j}$ with respect to $\CH_{[E-\Delta E, E]}$.
excluding $|\psi_{j}\rangle$ from $\CH_{[E-\Delta E, E]}$.
The next step is to consider if we can pick up another disentangled state in the orthogonal complement of $|\psi_{j}\rangle$ with respect to $\CH_{[E-\Delta E, E]}$. To answer this question, we will need to prove the following theorem. 

\begin{customthm}{3}\label{thm:Increment_bound}
Let $\CH$ be the Hilbert space of a bipartite system $\CH = \CH_A \otimes \CH_B$.
Let $\rho$ be a density matrix which can be written as
$\rho = \Pi_{\mathcal{E}}/W$,
where $\Pi_{\mathcal{E}}$ is the projection operator
onto a $W$-dimensional subspace $\mathcal{E} \cong \mathbb{C}^{W}$ of $\CH$.
Let $\{(p_i,|\psi_i\rangle)\}_{i=1,2,\cdots}$ be an ensemble of states which
generates $\rho$
% , i.e.,
% \begin{align}
%   \rho = \sum_i p_i |\psi_i\rangle\langle{\psi_i}|,
% \end{align}
with the average entanglement entropy between $A$ and $B$
\begin{align}
    \overline{E_{vN}} = \sum_i p_i E_{vN}(|\psi_i\rangle\langle{\psi_i}|). \label{eq:EvNbar_def}
\end{align}
Pick up an arbitrary set of $k$ $(< W/e^3$, where $e=2.718\dots$ is the Euler's number$)$ orthogonal states
$\{|\varphi_1\rangle,|\varphi_2\rangle,\cdots,|\varphi_k\rangle\} \subset \CE$
and consider its orthogonal complement 
\begin{align}
    \mathcal{E}' = \mathcal{E} \cap \left(\mathrm{span}\{|\varphi_1\rangle,|\varphi_2\rangle,\cdots,|\varphi_k\rangle\}\right)^\perp. \label{eq:Approx_subspace}
\end{align}
Let $\Pi_{\mathcal{E}}'$ denote the projection operator onto $\mathcal{E}'$.
Then there exists an ensemble of states $\{(p_i',|\psi_i'\rangle)\}_{i=1,2,\cdots}$
generating the density matrix $\rho' \propto \Pi_{\mathcal{E}'}$
% on $\mathcal{E}'$
whose average entanglement entropy 
\begin{align}
    \overline{E_{vN}}' = \sum_i p_i' E_{vN}(|\psi_i'\rangle\langle{\psi_i'}|). 
\end{align}
can be bounded from two sides as
\begin{align}
  &\overline{E_{vN}}'
  < \frac{1}{1-k/W} \left[ \overline{E_{vN}} + (k/W)^{1/3} \left\{\frac{1}{3} \log W/k + 5 \log\dim\mathcal{H}_B\right\}
  + k/W \log\dim\mathcal{H}_B \right], \label{eq:AvgEnt_bound}\\
  &\overline{E_{vN}}'
  > \frac{1}{1-k/W} \left[ (1-2k/W) \overline{E_{vN}} - (k/W)^{1/3} \left\{\frac{1}{3} \log W/k + 5 \log\dim\mathcal{H}_B\right\}
  - k/W \log\dim\mathcal{H}_B \right].
\end{align}
\end{customthm}

\begin{proof}
Let us show the theorem in a constructive way.
It is crucial to note that, since $|\varphi_i\rangle \in \CE$,
\begin{align}
  \sum_i p_i \langle{\psi_i|P|\psi_i}\rangle
  = \mathrm{Tr} \rho P
  = \mathrm{Tr} \left[ \frac{\Pi_{\mathcal{E}}}{W} \sum_{i=1}^k |\varphi_i\rangle\langle\varphi_i| \right]
  = k/W.\label{eq:AvgInnerProd_bound}
\end{align}
Here $P$ is the projection operator onto the subspace
spanned by the excluded states in \eqref{eq:Approx_subspace}.
This implies $\langle{\psi_i|P|\psi_i}\rangle$ is small with a large probability when $W$ is large, which motivates us to take
\begin{align}
  p_i' &= \frac{1-\langle{\psi_i|P|\psi_i}\rangle}{1-k/W} p_i, \label{eq:P_def}\\
  |{\psi_i'}\rangle &= \frac{(1-P)|\psi_i\rangle}{\sqrt{1-\langle{\psi_i|P|\psi_i}\rangle}}. \label{eq:Psi_def}
\end{align}
It can be easily checked that
this $\{(p_i',|\psi_i'\rangle)\}$ generates $\rho'$ as follows:
\begin{align}
  \Pi_{\mathcal{E}'}
  &= \Pi_{\mathcal{E}} - P
  = (\Pi_{\mathcal{E}} - P) \Pi_{\mathcal{E}} (\Pi_{\mathcal{E}} - P) \nonumber\\
  &= (\Pi_{\mathcal{E}} - P) \left[W\sum_i p_i |{\psi_i}\rangle\langle{\psi_i}|\right] (\Pi_{\mathcal{E}} - P)
  = (W-k) \sum_i p_i' |{\psi_i'}\rangle\langle{\psi_i'}|. 
\end{align}
% positivity of $\Pi_{\mathcal{E}}$ is evident since it can be written as a convex mixture of pure states
Then we proceed to consider how the average entanglement entropy changes
when moving from $\{(p_i,|\psi_i\rangle)\}$ to $\{(p_i',|\psi_i'\rangle)\}$.
Using the triangle inequality, we have
\begin{align}
  |\overline{E_{vN}}'-\overline{E_{vN}}|
  &= \left| \sum_i p_i' E_{vN}(|\psi_i'\rangle\langle{\psi_i'}|) - \sum_i p_i E_{vN}(|\psi_i\rangle\langle{\psi_i}|)) \right| \nonumber\\
  &= \left| \sum_i 
    (p_i'-p_i) E_{vN}(|\psi_i\rangle\langle{\psi_i}|)
    + \sum_i p_i' (E_{vN}(|\psi_i'\rangle\langle{\psi_i'}|)-E_{vN}(|\psi_i\rangle\langle{\psi_i}|))
   \right|\nonumber\\
  &\leq \underbrace{\sum_i |p_i'-p_i|
  E_{vN}(|\psi_i\rangle\langle{\psi_i}|)}_{=(\ast 1)}
  + \underbrace{\sum_i p_i' |E_{vN}(|\psi_i'\rangle\langle{\psi_i'}|)-E_{vN}(|\psi_i\rangle\langle{\psi_i}|)|}_{=(\ast 2)}.
\end{align}
First we give an upper bound for $(\ast 1)$.
By definition, we have
\begin{align}
  p_i'-p_i = \frac{k/W}{1-k/W} p_i - \frac{1}{1-k/W} p_i \langle\psi_i|P|\psi_i\rangle.
\end{align}
Substituting this into $(\ast 1)$, we obtain
\begin{align}
  (\ast 1)
  &\leq \frac{k/W}{1-k/W} \sum_i p_i E_{vN}(|\psi_i\rangle\langle{\psi_i}|)
  + \frac{1}{1-k/W} \sum_i p_i \langle\psi_i|P|\psi_i\rangle E_{vN}(|\psi_i\rangle\langle{\psi_i}|) \nonumber\\
  &\leq \frac{k/W}{1-k/W} \sum_i p_i E_{vN}(|\psi_i\rangle\langle{\psi_i}|)
  + \frac{1}{1-k/W} \log\dim\mathcal{H}_B \sum_i p_i \langle\psi_i|P|\psi_i\rangle \nonumber\\
  &= \frac{k/W}{1-k/W} \{ \overline{E_{vN}} + \log\dim\mathcal{H}_B \}.
\end{align}
Here we have used Eqs.~\eqref{eq:EvNbar_def} and \eqref{eq:AvgInnerProd_bound}.

As the next step, we consider bounding $(\ast 2)$ from above. It would be useful to note the following facts: 
\begin{itemize}
    \item Since for any state $|\phi\rangle\in\CH$, $0\leq E_{vN}(|\phi\rangle\langle{\phi}|)\leq \log\dim\mathcal{H}_B$, we have 
            \begin{align}\label{eq:bound1}
                |E_{vN}(|\psi_i'\rangle\langle{\psi_i'}|)-E_{vN}(|\psi_i\rangle\langle{\psi_i}|)|
               \leq \log\dim\mathcal{H}_B.
            \end{align}
    
    \item Using Eq.~\eqref{eq:Psi_def}, we have
            \begin{align}
                \|\mathrm{Tr}_A|{\psi_i'}\rangle\langle{\psi_i'}|-\mathrm{Tr}_A|{\psi_i}\rangle\langle{\psi_i}|\|_1
                \leq \||{\psi_i'}\rangle\langle{\psi_i'}|-|{\psi_i}\rangle\langle{\psi_i}|\|_1
                = 2 \sqrt{1-|\langle{\psi_i'|\psi_i}\rangle|^2}
                = \sqrt{4\langle\psi_i|P|\psi_i\rangle}.
            \end{align}
            Therefore, by applying Fannes' inequality, we find
            \begin{align}
                |E_{vN}(|\psi_i'\rangle\langle{\psi_i'}|)-E_{vN}(|\psi_i\rangle\langle{\psi_i}|)|
                &\leq \sqrt{4\langle\psi_i|P|\psi_i\rangle} \log\dim\mathcal{H}_B
                - \sqrt{4\langle\psi_i|P|\psi_i\rangle} \log \sqrt{4\langle\psi_i|P|\psi_i\rangle},
            \end{align}
            for $i$ such that $\sqrt{4\langle\psi_i|P|\psi_i\rangle} \leq 1/e$.
            This gives a better bound compared to \eqref{eq:bound1} when $\langle\psi_i|P|\psi_i\rangle$ is sufficiently small.
            Since we have assumed $k < W/e^3$, this inequality holds for any $i$ such that $4\langle\psi_i|P|\psi_i\rangle \leq (k/W)^{2/3}$.

    \item   % For any $\epsilon > 0$,
            One finds
            \begin{align}
                \sum_{\substack{i\ \mathrm{s.t.}\\4\langle\psi_i|P|\psi_i\rangle > (k/W)^{2/3}}} p_i
                < 4 (k/W)^{1/3}. \label{eq:EntangledProb_bound}
            \end{align}
            This can be seen as follows.
            \begin{align}
                (k/W)^{2/3} \sum_{\substack{i\ \mathrm{s.t.}\\4\langle\psi_i|P|\psi_i\rangle > (k/W)^{2/3}}} p_i
                < 4 \sum_{\substack{i\ \mathrm{s.t.}\\4\langle\psi_i|P|\psi_i\rangle > (k/W)^{2/3}}} p_i \langle\psi_i|P|\psi_i\rangle
                \leq 4 \sum_i p_i \langle\psi_i|P|\psi_i\rangle
                = 4 k/W,
            \end{align}
            where we used \eqref{eq:AvgInnerProd_bound}. 
\end{itemize}
Based on these, it is straightforward to see
\begin{align}
  (\ast 2)
  &\leq \frac{1}{1-k/W} \sum_i p_i |E_{vN}(|\psi_i'\rangle\langle{\psi_i'}|)-E_{vN}(|\psi_i\rangle\langle{\psi_i}|)| \label{eq:first_line}\\
  &= \frac{1}{1-k/W} \sum_{\substack{i\ \mathrm{s.t.}\\4\langle\psi_i|P|\psi_i\rangle > (k/W)^{2/3}}} p_i |E_{vN}(|\psi_i'\rangle\langle{\psi_i'}|)-E_{vN}(|\psi_i\rangle\langle{\psi_i}|)|\nonumber\\
  &~~~+ \frac{1}{1-k/W} \sum_{\substack{i\ \mathrm{s.t.}\\4\langle\psi_i|P|\psi_i\rangle \leq (k/W)^{2/3}}} p_i |E_{vN}(|\psi_i'\rangle\langle{\psi_i'}|)-E_{vN}(|\psi_i\rangle\langle{\psi_i}|)|\\
  &\leq \frac{1}{1-k/W} \log\dim\mathcal{H}_B \sum_{\substack{i\ \mathrm{s.t.}\\4\langle\psi_i|P|\psi_i\rangle > (k/W)^{2/3}}} p_i\nonumber\\
  &~~~+ \frac{1}{1-k/W} \{(k/W)^{1/3} \log\dim\mathcal{H}_B - (k/W)^{1/3} \log (k/W)^{1/3}\} \sum_{\substack{i\ \mathrm{s.t.}\\4\langle\psi_i|P|\psi_i\rangle \leq (k/W)^{2/3}}} p_i\\
  &< \frac{(k/W)^{1/3}}{1-k/W} \left\{\frac{1}{3} \log W/k + 5 \log\dim\mathcal{H}_B\right\},
\end{align}
where we used \eqref{eq:P_def} in the first line and \eqref{eq:EntangledProb_bound} in the last line.
% By taking
% \begin{align}
%   \epsilon = \frac{1}{3} \left(1-\frac{\log k}{\log W}\right),
% \end{align}
% 
Combining $(\ast 1)$ and $(\ast 2)$, we obtain
\begin{align}
  |\overline{E_{vN}}'-\overline{E_{vN}}|
  < \frac{(k/W)^{1/3}}{1-k/W} \left\{\frac{1}{3} \log W/k + 5 \log\dim\mathcal{H}_B\right\}
  + \frac{k/W}{1-k/W} \{ \overline{E_{vN}} + \log\dim\mathcal{H}_B \}.
\end{align}
This gives both the upper bound and the lower bound shown in the theorem. 
\end{proof}

\subsection{Existence of \texorpdfstring{$e^{S_{\rm micro}}$}{exp[Smicro]} disentangled microstates}\label{sec:hol_final_statement}

In this subsection, we combine the results presented in the previous subsections to get our final claim.
As the first step, let us consider the microcanonical ensemble $\rho^{\rm (micro)}$ as $\rho$, the associated microcanonical subspace 
\begin{align}
    \CH_{[E-\Delta E, E]} = \mathrm{span} \left\{ |E_i\rangle \in \mathcal{H} \middle| E_i \in [E-\Delta E, E] \right\},
\end{align}
as $\CE$, and a disentangled state $|\psi_j\rangle$ appearing in an ensemble of states in proposition \ref{prop:hol_bound} as $|\varphi_1\rangle$ (which is guaranteed by proposition \ref{lemma}) in theorem \ref{thm:Increment_bound}. Let us consider the $k=1$ case of theorem \ref{thm:Increment_bound}.
Then $W \equiv e^{S_{\rm micro}} \sim e^{S_{BH}} = e^{A_{\rm horizon}/4G_N}$, where $A_{\rm horizon}$ is the horizon area of the AdS black hole corresponding to $\rho^{\rm (micro)}$. In this case, since $\overline{E_{vN}} = O(1/G_N) = O({S_{BH}}) = O({S_{\rm micro}})$ and $\log\dim\CH_B = O(1/G_N)= O({S_{BH}}) = O({S_{\rm micro}})$, according to the upper bound \eqref{eq:AvgEnt_bound} given in theorem \ref{thm:Increment_bound}, $\rho' \propto \Pi_{\mathcal{E}'}$ can be generated by an ensemble of states whose average entanglement entropy is changed by at most $O(e^{-S_{\rm micro}/3}{S_{\rm micro}})$ from $\overline{E_{vN}}$.
%Then $W\sim e^{S_{BH}} = e^{A_{\rm horizon}/4G_N}$, where $A_{\rm horizon}$ is the horizon area of the AdS black hole corresponding to $\rho^{\rm (micro)}$. In this case, since $\overline{E_{vN}} = O(1/G_N) = O({S_{BH}})$ and $\log\dim\CH_B = O(1/G_N)= O({S_{BH}})$, according to the upper bound \eqref{eq:AvgEnt_bound} given in theorem \ref{thm:Increment_bound}, $\rho' \propto \Pi_{\mathcal{E}'}$ can be generated by an ensemble of states whose average entanglement entropy is changed by at most $O(e^{-S_{BH}/3}{S_{BH}})$ from $\overline{E_{vN}}$.

It should be mentioned that, however, $1/G_N$ is not the only parameter involved in this order estimation. Since we are considering a CFT, where the entanglement entropy is not well-defined, we need to perform a lattice regularization. This procedure introduces a UV cutoff $\epsilon$, where $\log\dim\mathcal{H}_B$ is $O(1/\epsilon^d)$ and hence diverges in the continuum limit $\epsilon \to 0$. Throughout this paper, we assume the existence of a lattice regularization of the CFT and discuss the scaling behavior against $1/G_N$ with fixed $\epsilon$. The arguments presented here are valid in this regime. On the other hand, our arguments will not be valid if one takes the continuum limit before taking other limits.

Then we can take another disentangled state $|\psi'_{j'}\rangle$ from the ensemble that generates $\rho'$ obtained in the first step, and regard it as $|\varphi_2\rangle$. Applying the $k=2$ case of theorem \ref{thm:Increment_bound}, we can find an ensemble of states generating the microcanonical ensemble associated with
\begin{align}
    \CH_{[E-\Delta E, E]} \cap \left(\mathrm{span}\{|\varphi_1\rangle,|\varphi_2\rangle\}\right)^\perp,
\end{align}
with average entanglement entropy changed by at most $O(e^{-S_{\rm micro}/3}{S_{\rm micro}})$ from $\overline{E_{vN}}$. 

% According to theorem \ref{thm:Increment_bound}, one can repeat this procedure $k$ times without changing the average entanglement entropy from $\overline{E_{vN}}$ in the leading order in $1/G_N$, provided that $k$ satisfies
% \begin{align}
%     \log k \leq S_\mathrm{micro} - 3 \left\{ \log\log\dim\mathcal{H}_B - \log\overline{E_{vN}} \right\} - f(1/G_N), \label{eq:k_bound}
% \end{align}
% for some function $f(x)$ that diverges as $x\to\infty$, such as $\log x$, $\log\log x$ and so on. Therefore one can obtain $\exp\left[S_{\rm micro}-f(S_{\rm micro})\right]$ disentangled states as
According to theorem \ref{thm:Increment_bound}, one can repeat this procedure for at least $\exp\left[S_{\rm micro}-f(S_{\rm micro})\right]$ times, where $f(x)$ is any function which diverges at $x\rightarrow \infty$ such as $\log x$, $\log\log x$ and so on, without changing the average entanglement entropy in the leading order, i.e. $O(1/G_N)$. Therefore one can obtain $\exp\left[S_{\rm micro}-f(S_{\rm micro})\right]$ disentangled states as
\begin{align}
  |\varphi_1\rangle, |\varphi_2\rangle, |\varphi_3\rangle, \cdots.
\end{align}
By construction, these states are orthogonal.
To summarize, we obtain the following statement.
\begin{customcoro}{4}\label{coro:hol_final}
    In a holographic CFT, let $\CH_{[E-\Delta E, E]}$ be the microcanonical subspace whose dimension is given by $e^{S_{\rm micro}}$. There exists an orthogonal basis of $\CH_{[E-\Delta E, E]}$ such that at least $\exp\left[S_{\rm micro} - f(S_{\rm micro})\right]$ elements are disentangled states, for any $f(x) \in \omega(1)$.
\end{customcoro}
Although we made our statement in terms of the entropy computed from the microcanonical ensemble, the disentangled states we found are sufficient to give the leading order of the Bekenstein-Hawking entropy $S_{BH}$, since $S_{BH}$ and $S_{\rm micro}$ matches at the leading order.

Note that we have not only proven the existence of a basis satisfying requirements mentioned in corollary \ref{coro:hol_final}, but also given an algorithm to construct such a basis from an ensemble of states satisfying requirements mentioned in proposition \ref{prop:hol_bound} in the proof of theorem \ref{thm:Increment_bound}. We will come back to discuss if one can give an ensemble of states satisfying requirements mentioned in proposition \ref{prop:hol_bound} in section \ref{sec:constructing_disent}. Besides, we also note that if one only wants to prove the existence of $\exp\left[S_{\rm micro}-f(S_{\rm micro})\right]~(\forall f(x)\in\omega(1))$ disentangled states but does not care about how to construct them, there actually exists a more simple proof. We will leave it in appendix \ref{app:proof}.

\section{Area law states in quantum many-body systems} \label{sec:thermodynamic_limit}

In section \ref{sec:hol_CFT},
we discussed disentangled states in holographic CFT
at the semiclassical limit $G_N \rightarrow 0$.
In this section, we see that we can get a similar result
for generic short-ranged Hamiltonian with $L \gg 1$.
Here, $L$ is the linear scale of the system.
Accordingly, the volume of the system is $O(L^d)$,
where $d$ is the system dimension. 
Again, let us divide the system into two parts $A$ and $B$,
where $B$ is slightly smaller than $A$.
A typical state in the microcanonical subspace $\CH_{[E-\Delta E,E]}$ would be a volume law state,
i.e. its entanglement entropy would be $O(L^d)$.
Below in this section we show that there exists an orthogonal basis of the microcanonical subspace $\CH_{[E-\Delta E, E]}$ such that at least $\exp\left[S_{\rm micro}-O(\log S_{\rm micro})\right]$ elements are area law states, where $S_{\rm micro} = \log \dim \CH_{[E-\Delta E,E]}$ is the thermodynamic entropy computed from the microcanonical ensemble.

We take a similar (but slightly different) strategy as in section \ref{sec:hol_CFT}.
We will firstly make a conjecture\footnote{It is worth noting that holographic CFTs are explicit examples where this conjecture can be shown as we will see later.} about the entanglement entropy in microcanonical TFD states and give an upper bound to the entanglement of formation of the microcanonical ensemble $\rho^{\rm (micro)}$ under the conjecture.
Then we can apply proposition \ref{lemma} and theorem \ref{thm:Increment_bound} in a similar way as in section \ref{sec:hol_final_statement} to show there exist at least $\exp\left[S_{\rm micro}-O(\log S_{\rm micro})\right]$ orthogonal area law states in $\CH_{[E-\Delta E,E]}$.

\subsection{Conjecture for \texorpdfstring{$E_{RE}$}{the reflected entropy}}
We would like to first give an upper bound to the entanglement of formation of the microcanonical ensemble $\rho^{\rm (micro)}$.
Different from the holographic case, now the entanglement of purification is not accessible.
However, since we are now looking at the scaling behavior with respect to $L$,
we can give it a looser upper bound using the reflected entropy.
The canonical purification of the microcanonical ensemble
is simply given by the microcanonical TFD state defined in \eqref{eq:MCTFD}.
Both the canonical ensemble and microcanonical ensemble
are statistical mixtures of energy eigenstates,
whose probability distribution has a sharp peak
around the energy expectation value in the thermodynamic limit ($L\rightarrow\infty$).
Therefore, for sufficiently large $L$, their canonical purifications are expected to have similar entanglement properties.
In the canonical TFD states \eqref{eq:TFD},
the entanglement entropy is extensively studied for various systems  \cite{Barthel2017,HM13}.
In particular for $d=1$, it has been shown that the canonical TFD state obeys an area law by the field-theoretical analysis \cite{Barthel2017}
and rigorously proven for sufficiently high temperature \cite{Kuwahara2021}. Besides, this has also been shown in holographic CFTs from gravitational computations \cite{HM13}.
From our knowledge about TFD states, it is plausible to make the following conjecture.

\begin{customconj}{5}\label{conjecture}
    Consider a $d$-dimensional quantum many-body system with local interactions,
    and divide the spatial region into $A$ and $B$ where $B$ is slightly smaller than $A$.
    Consider a microcanonical TFD state defined on $\CH_A \otimes \CH_B \otimes \CH_{A^*} \otimes \CH_{B^*}$.
    Then the entanglement entropy between $AA^*$ and $BB^*$ is $O(L^{d-1})$,
    i.e. it obeys the area law. 
\end{customconj}

It is worth noting one can show conjecture \ref{conjecture} is true in holographic CFTs. First, according to \cite{Marolf18}, microcanonical TFD states are also geometric states corresponding to two-sided eternal black holes and the RT formula hold in them. Then one can perform the same computation in \cite{HM13} to find that the entanglement entropy indeed obeys an area law.

\subsection{Existence of \texorpdfstring{$e^{S_{\rm micro}}$}{exp[Smicro]} area law microstates}\label{sec:qmb_final_statement}

Combining this conjecture and the general relation between entanglement of formation and reflected entropy \eqref{eq:F_P_RE}, we have
\begin{align}
  E_F \leq E_{RE} = O(L^{d-1}).
\end{align}
Therefore, by definition of $E_F$,
it is straightforward to say that there exists an ensemble of states generating $\rho^{\rm (micro)}$
such that the averaged entanglement entropy is $O(L^{d-1})$.
Thus, there must exist at least one area law state $|\varphi_1\rangle$ in the microcanonical subspace according to proposition \ref{lemma}.
Then we proceed to consider if we can pick up another disentangled state in the orthogonal complement of $|\varphi_1\rangle$. 
The discussion here is parallel with that in section \ref{sec:constructing_disent}. 

As the first step, consider the microcanonical ensemble $\rho^{\rm (micro)}$ as $\rho$, the associated microcanonical subspace $\CH_{[E-\Delta E, E]}$ as $\CE$, and an area law state appearing in an ensemble of states, whose average entanglement entropy is $O(L^{d-1})$, as $|\varphi_1\rangle$ in theorem \ref{thm:Increment_bound}. Again, let us first consider the $k=1$ case of theorem \ref{thm:Increment_bound}.
Then $W \equiv e^{S_{\rm micro}} = e^{O(L^d)}$. In this case, since $\overline{E_{vN}} = O(L^{d-1}) = O({S}_{\rm micro}^{\frac{d-1}{d}})$ and $\log\dim\CH_B = O(L^d) = O({S_{\rm micro}})$, according to the upper bound \eqref{eq:AvgEnt_bound} given in theorem \ref{thm:Increment_bound}, $\rho' \propto \Pi_{\mathcal{E}'}$ can be generated by an ensemble of states whose average entanglement entropy is changed by at most $O(e^{-S_{\rm micro}/3}{S_{\rm micro}})$ from $\overline{E_{vN}}$.

Then we can take another area law state from the ensemble that generates $\rho'$ obtained in the first step as $|\varphi_2\rangle$. Again, applying the $k=2$ case of theorem \ref{thm:Increment_bound}, we can find an ensemble of states generating the microcanonical ensemble associated with
\begin{align}
    \CH_{[E-\Delta E, E]} \cap \left(\mathrm{span}\{|\varphi_1\rangle,|\varphi_2\rangle\}\right)^\perp,
\end{align}
with average entanglement entropy changed by at most $O(e^{-S_{\rm micro}/3}{S_{\rm micro}})$ from the original $\overline{E_{vN}}$. From theorem \ref{thm:Increment_bound}, one can repeat this procedure for at least $\exp\left[S_{\rm micro}-O(\log S_{\rm micro})\right]$ times to get orthogonal area law states in $\CH_{[E-\Delta E, E]}$. As a result, we have the following corollary. 
\begin{customcoro}{6}\label{coro:qmb}
    In a $d$-dimensional quantum many-body system with local interactions,
    let $\CH_{[E-\Delta E, E]}$ be the microcanonical subspace whose dimension is given by $e^{S_{\rm micro}}$,
    and divide the spatial region into $A$ and $B$ where $B$ is slightly smaller than $A$.
    Assuming conjecture \ref{conjecture},
    then there exists an orthogonal basis of $\CH_{[E-\Delta E, E]}$ such that at least $\exp\left[S_{\rm micro}-O(\log S_{\rm micro})\right]$ elements are area law states. 
\end{customcoro}
Although we have discussed in terms of $S_{\rm micro}$, the area law states we found are sufficient to give the leading order of the thermodynamic entropy computed from the canonical ensemble $S_{\rm can}$, since $S_{\rm can}$ and $S_{\rm micro}$ matches at the leading order.

Here, we have applied the cumbersome theorem \ref{thm:Increment_bound} to prove corollary \ref{coro:qmb}. In fact, if combined with another theorem, this proof can even let us construct an explicit example of such $\exp\left[S_{\rm micro}-O(\log S_{\rm micro})\right]$ orthogonal area law states, as we will see in section \ref{sec:construction}. However, if one merely wants to prove corollary \ref{coro:qmb} without giving any explicit construction, there actually exists a more straightforward way. We will present such an alternative proof in appendix \ref{app:proof}.

\section{Constructing area law states}\label{sec:construction}

In section \ref{sec:thermodynamic_limit},
we have shown that there exists an orthogonal basis of the microcanonical subspace $\CH_{[E-\Delta E, E]}$
such that at least $\exp\left[S_{\rm micro}-O(\log S_{\rm micro})\right]$ elements are area law states under a plausible conjecture.
In this section, we provide a concrete method
for constructing $\exp\left[S_{\rm micro}-O(\log S_{\rm micro})\right]$ area law states.

In section \ref{sec:construction_ensemble},
we present an explicit example of an ensemble of states
generating the microcanonical ensemble whose average entanglement entropy is bounded by the reflected entropy. Therefore, according to conjecture \ref{conjecture}, the average entanglement entropy is at most $O(L^{d-1})$.
General results got in this section will be summarized as a theorem.
Then in section \ref{sec:construction_arealawstates},
we will show that we can construct at least $\exp\left[S_{\rm micro}-O(\log S_{\rm micro})\right]$ orthogonal area law states
from the ensemble of states constructed in section \ref{sec:construction_ensemble}
using the algorithm presented in the proof of theorem \ref{thm:Increment_bound}.

\subsection{Constructing an ensemble of states with low entanglement}\label{sec:construction_ensemble}
In this subsection, we construct an ensemble of states $\{(p_i,|\psi_i\rangle)\}_{i=1,2,\cdots}$ that generates $\rho^{\rm (micro)}$
and 
% give an upper bound for the average entanglement entropy using the conjecture \ref{conjecture} for the reflected entropy of $\rho^{\rm (micro)}$
show that the average entanglement entropy associated with this ensemble is bounded by the reflected entropy and hence follows an area law under conjecture \ref{conjecture}.

To be concrete, let us consider a discrete system defined on a lattice with $L^d$ sites. For a continuous quantum field theory, we can consider its lattice regularization. Then we have $O(e^{L^d})$ orthogonal product states which span the Hilbert space of the whole system. Let us denote the set of these product states as $\{|P_i\rangle\}_{i=1,2,\cdots}$. Since the microcanonical ensemble $\rho^{\rm (micro)}$ can be decomposed as 
\begin{align}
    \rho^{\rm (micro)} &= e^{-S_{\rm micro}} \Pi_{[E-\Delta E, E]} \nonumber\\
    &= e^{-S_{\rm micro}} \Pi_{[E-\Delta E, E]} \left(\sum_{i}|P_i\rangle\langle P_i|\right)  \Pi_{[E-\Delta E, E]} \nonumber\\
    &= \sum_{i} \left(e^{-S_{\rm micro}/2} \Pi_{[E-\Delta E, E]} |P_i\rangle\right) \left(\langle P_i| \Pi_{[E-\Delta E, E]}e^{-S_{\rm micro}/2}\right) \nonumber\\
    &= \sum_{i} e^{-S_{\rm micro}}\langle P_i |\Pi_{[E-\Delta E, E]}| P_i \rangle
    \left(\frac{ \Pi_{[E-\Delta E, E]} |P_i\rangle}{\sqrt{\langle P_i |\Pi_{[E-\Delta E, E]}| P_i \rangle}}\right) \left(\frac{\langle P_i| \Pi_{[E-\Delta E, E]}}{\sqrt{\langle P_i |\Pi_{[E-\Delta E, E]}| P_i \rangle}}\right), 
\end{align}
it seems plausible to expect that 
\begin{align}
    p_i &= e^{-S_{\rm micro}}\langle P_i |\Pi_{[E-\Delta E, E]}| P_i \rangle, \\
    |\psi_i\rangle &= \frac{ \Pi_{[E-\Delta E, E]} |P_i\rangle}{\sqrt{\langle P_i |\Pi_{[E-\Delta E, E]}| P_i, \rangle}}, \label{eq:MCMETTS}
\end{align}
gives a small average entanglement. This is indeed the case that the average entanglement entropy in this case can be bounded by the reflected entropy of $\rho^{\rm (micro)}$. In fact, we can show a more general theorem presented below.

\begin{customthm}{7} \label{thm:METTS_RE}
Let $\rho$ be a density matrix defined on a bipartite system $\CH = \CH_A \otimes \CH_B$.
Take any set of orthogonal states $\{|P_i\rangle_{A(B)}\}$
for which $\sum_i |P_i \rangle\langle P_i|_{A(B)}$ gives the unity matrix $I_{A(B)}$ on $\CH_{A(B)}$.
Then, using product states $|P_{ij}\rangle = |P_i\rangle_A \otimes |P_j\rangle_B$,
one can decompose $\rho$ as
\begin{align}
    \rho
    = \sum_{i,j} \sqrt{\rho} |P_{ij} \rangle \langle P_{ij}| \sqrt{\rho}
    = \sum_{i,j} p_{ij} |\psi_{ij}\rangle \langle\psi_{ij}|,
\end{align}
where 
\begin{align}
    p_{ij} &= \langle P_{ij} |\rho| P_{ij} \rangle,\\
    |\psi_{ij}\rangle &= \frac{\sqrt{\rho} |P_{ij}\rangle}{\sqrt{\langle P_{ij} |\rho| P_{ij} \rangle}}.
\end{align}
For this ensemble of states, the average entanglement entropy is upper bounded by the reflected entropy of $\rho$:
% In this case, if any $|P_i\rangle$ is a product state between $A$ and $B$, i.e. $E_{vN}(|P_i\rangle\langle P_i|) = 0$, then
\begin{align}\label{eq:upperbound_vN}
  \overline{E_{vN}} = \sum_{i,j} p_{ij} E_{vN}\left(|\psi_{ij}\rangle\langle \psi_{ij}|\right)
  \leq E_{RE}(\rho),
\end{align}
% i.e. the average entanglement entropy is upper bounded by the reflected entropy. 
if $|P_{ij}^*\rangle = |P_i^*\rangle_A \otimes |P_j^*\rangle_B$.\footnote{Note that this assumption holds for any theory whose CPT transformation respects the locality. This is the type of theory we are interested in.}
\end{customthm}

\begin{proof}
A key observation in our proof is that
\begin{align}
  I_{A} \otimes I_{B} \otimes |P_i^* \rangle\langle P_i^*|_{A^*} \otimes |P_j^* \rangle\langle P_j^*|_{B^*} |\sqrt{\rho}\rangle &= \sqrt{p_{ij}} |\psi_{ij}\rangle |P_{ij}^* \rangle, \label{eq:TFD-METTS}
\end{align}
where $|\sqrt{\rho}\rangle$ is the canonical purification of $\rho$.
Indeed, by definition of the canonical purification \eqref{eq:can_purification}, we have
\begin{align}
    |\sqrt{\rho}\rangle
    &= \sqrt{\rho} \otimes I \sum_\nu |\varphi_\nu\rangle |\varphi_\nu^*\rangle
    = \sqrt{\rho} \otimes I \sum_\nu \left( \sum_{i,j}| P_{ij} \rangle \langle P_{ij} | \right) |\varphi_\nu\rangle |\varphi_\nu^*\rangle \nonumber\\
    &= \sqrt{\rho} \otimes I \sum_\nu \sum_{i,j} \langle P_{ij} | \varphi_\nu \rangle | P_{ij} \rangle |\varphi_\nu^*\rangle
    = \sqrt{\rho} \otimes I \sum_\nu \sum_{i,j} \langle \varphi_\nu^* | P_{ij}^* \rangle | P_{ij} \rangle |\varphi_\nu^*\rangle \nonumber\\
    &= \sqrt{\rho} \otimes I \sum_{i,j} | P_{ij} \rangle \left( \sum_\nu |\varphi_\nu^*\rangle \langle\varphi_\nu^*| \right) |P_{ij}^*\rangle
    = \sqrt{\rho} \otimes I \sum_{i,j} | P_{ij} \rangle |P_{ij}^*\rangle
    = \sum_{i,j} \sqrt{p_{ij}} | \psi_{ij} \rangle |P_{ij}^*\rangle,
\end{align}
where in the second line, we have used the antiunitarity of the CPT operator.
Eq.~\eqref{eq:TFD-METTS} motivates us to evaluate the changes in the entanglement entropy
due to the local projective measurements on $A^*$ and $B^*$.

Consider the projection
\begin{align}
  K_j &= I_{A} \otimes I_{B} \otimes I_{A^*} \otimes |P_j^* \rangle\langle P_j^*|_{B^*}.
\end{align}
From the completeness of $K_j$, we get
\begin{align}
  &\mathrm{Tr}_{B B^*} |\sqrt{\rho}\rangle\langle\sqrt{\rho}|
  = \sum_j \mathrm{Tr}_{B B^*} \left[ K_j |\sqrt{\rho}\rangle\langle\sqrt{\rho}| {K_j}^\dagger \right].
\end{align}
% $\sum_j {K_j}^\dagger K_j = 1_{B B^*}$
Therefore, we have
\begin{align}
  E_{RE}(\rho)
  = S(\mathrm{Tr}_{B B^*} |\sqrt{\rho}\rangle\langle\sqrt{\rho}|)
  = S \left( \sum_j s_j \mathrm{Tr}_{B B^*} | K_j\sqrt{\rho}\rangle \langle K_j\sqrt{\rho}| \right),
\end{align}
where
\begin{align}
  |K_j\sqrt{\rho}\rangle &= \frac{K_j|\sqrt{\rho}\rangle}{\sqrt{\langle\sqrt{\rho}|K_j|\sqrt{\rho}\rangle}},\\
  s_j &= \langle\sqrt{\rho}|K_j|\sqrt{\rho}\rangle.
\end{align}
Further, using the matrix concavity of the von Neumann entropy,
we obtain
\begin{align}\label{eq:concavity}
  E_{RE}(\rho)
  \geq \sum_i s_j S(\mathrm{Tr}_{B B^*} |K_j\sqrt{\rho} \rangle\langle K_j\sqrt{\rho}|).
\end{align}
In the same way, we also obtain
\begin{align}
  &S(\mathrm{Tr}_{B B^*} |K_j\sqrt{\rho} \rangle\langle K_j\sqrt{\rho}|) \nonumber\\
  = &S(\mathrm{Tr}_{A A^*} |K_j\sqrt{\rho} \rangle\langle K_j\sqrt{\rho}|) \nonumber\\
  \geq& \sum_j r_{ij} S(\mathrm{Tr}_{A A^*} |J_i K_j\sqrt{\rho} \rangle\langle J_i K_j\sqrt{\rho}|) \nonumber\\
  = &\sum_j r_{ij} S(\mathrm{Tr}_{B B^*} |J_i K_j\sqrt{\rho} \rangle\langle J_i K_j\sqrt{\rho}|),
\end{align}
where
\begin{align}
  J_i &= I_{A} \otimes I_{B} \otimes |P_i^* \rangle\langle P_i^*|_{A^*} \otimes I_{B^*},\\
  |J_i K_j\sqrt{\rho}\rangle &= \frac{J_i K_j|\sqrt{\rho}\rangle}{\sqrt{\langle\sqrt{\rho}|J_i K_j|\sqrt{\rho}\rangle}},\\
  r_{ij} &= \frac{\langle\sqrt{\rho}|J_i K_j|\sqrt{\rho}\rangle}{\langle\sqrt{\rho}|K_j|\sqrt{\rho}\rangle}.
\end{align}
Thus, we have
\begin{align}
  E_{RE}(\rho)
  \geq \sum_{i, j} r_{ij} s_j S( \mathrm{Tr}_{B B^*} |J_i K_j\sqrt{\rho} \rangle\langle J_i K_j\sqrt{\rho}|).
\end{align}
Using Eq.~\eqref{eq:TFD-METTS}, we immediately see that
\begin{align}
    r_{ij} s_j &= p_{ij},\\
    | J_i K_j\sqrt{\rho}\rangle &= |\psi_{ij}\rangle |P_{ij}^* \rangle.
\end{align}
Then, from additivity and non-negativity of the von Neumann entropy, we finally obtain
\begin{align}
  E_{RE}(\rho)
  &\geq \sum_{i,j} p_{ij} \left\{ S(\mathrm{Tr}_{B} |\psi_{ij}\rangle \langle\psi_{ij}|) + S(\mathrm{Tr}_{B^*} |P_{ij}^* \rangle\langle P_{ij}^*|) \right\}
  \geq \sum_{i,j} p_{ij} E_{vN}(|\psi_{ij}\rangle\langle\psi_{ij}|)
  = \overline{E_{vN}}.
\end{align}    
Hence Eq.~\eqref{eq:upperbound_vN} is shown.
\end{proof}

\subsection{Constructing \texorpdfstring{$e^{S_{\rm micro}}$}{exp[S]} area law states}\label{sec:construction_arealawstates}

Proposition \ref{lemma} and theorem \ref{thm:Increment_bound} are universal for any quantum system
since no properties of specific physical systems are used in their proofs.
Therefore, we can follow the algorithm presented in the proof of theorem \ref{thm:Increment_bound} to construct orthogonal area law states. 
Starting from the ensemble of states \eqref{eq:MCMETTS}, we can pick up one area law state from them, according to theorem \ref{thm:METTS_RE} and conjecture \ref{conjecture}. 
Then we can construct another orthogonal area law state with the procedure presented in the proof of theorem \ref{thm:Increment_bound}. 
Again, we can repeat this procedure for at least $\exp\left[S_{\rm micro}-O(\log S_{\rm micro})\right]$ times
without changing area law scaling of the average entanglement entropy. 

Following this procedure, one can explicitly construct $\exp\left[S_{\rm micro}-O(\log S_{\rm micro})\right]$ orthogonal area law states from the states given in \eqref{eq:MCMETTS}.

\subsection{Constructing \texorpdfstring{$e^{S_{\rm micro}}$}{exp[SBH]} disentangled states in holographic CFTs?}\label{sec:constructing_disent}

One may wonder if the state ensemble in \eqref{eq:MCMETTS} provides an example satisfying requirements in proposition \ref{prop:hol_bound}, from which we can construct exponentially many orthogonal disentangled states in holographic CFTs. 

For this purpose, theorem \ref{thm:METTS_RE} turns out to be useless, since in a holographic theory, $E_{RE} = 2E_P$ \cite{DF19} and the difference between them $E_{RE} - E_P$ is $O(1/G_N) = O(S_{BH}) = O(S_{\rm micro})$, i.e. at the leading order of $1/G_N$. On the other hand, numerical results in various one-dimensional critial quantum spin chains (including both integrable ones and chaotic ones) \cite{KTWY23} show that $\overline{E_{vN}}$ over the ensemble \eqref{eq:MCMETTS} is given by $E_{RE}/2$. This indicates that we may expect \eqref{eq:MCMETTS} satisfies requirements in proposition \ref{prop:hol_bound}. We leave confirming $\overline{E_{vN}} \approx E_{RE}/2$ in more general dimensions and systems (hopefully even in holographic CFTs) as a future problem.

\section{Summary and Discussions}\label{sec:conclu}

In this paper, we have studied two types of low-entanglement atypical states in holographic CFTs and generic quantum many-body systems. Our results can be summarized as follows: 
\begin{itemize}
    \item There exist at least $\exp\left[S_{\rm micro} - f(S_{\rm micro})\right]$ ($\forall f(x)\in\omega(1)$) orthogonal disentangled black hole microstates in the microcanonical subspace $\CH_{[E-\Delta E,E]}$ in holographic CFTs at the semiclassical limit $G_N\rightarrow0$.
    \item There exist at least $\exp\left[S_{\rm micro}-O(\log S_{\rm micro})\right]$ orthogonal area law states in the microcanonical subspace in generic quantum many-body systems, under the assumption that the microcanonical TFD obeys an area law of entanglement. Holographic CFTs and generic one-dimensional quantum many-body systems with short-ranged interactions satisfy this assumption. 
    \item We have presented an explicit algorithm to construct $\exp\left[S_{\rm micro} - f(S_{\rm micro})\right]$ orthogonal area law states in generic quantum many-body systems and discussed its potential to be used to construct disentangled states in holographic CFTs. 
\end{itemize}
Let us discuss possible improvements of our arguments, relations with other topics including additivity conjectures and black hole microstate counting, and future directions below. 

\subsection{Connections to additivity conjectures}
In \cite{HP20}, the authors argue that the nonexistence of extensive violations of additivity conjectures leads to the existence $\exp\left[S_{\rm can}-O(1)\right]$ orthogonal disentangled microstates. Our results, the existence $\exp\left[S_{\rm micro}-f(S_{\rm micro})\right]$ ($\forall f(x) \in \omega(1)$) orthogonal disentangled microstates, is not directly comparable with this since the entropies used are computated from different ensembles.
However, we would like to note that since $S_{\rm micro}$ and $S_{\rm can}$ matches at the leading order, the associated physics is essentially the same. Moreover, we can even argue that the existence $\exp\left[S_{\rm micro}-f(S_{\rm micro})\right]$ ($\forall f(x) \in \omega(1)$) orthogonal disentangled microstates implies the existence $\exp\left[S_{\rm can}-O(1)\right]$ orthogonal disentangled microstates, under some assumptions from thermodynamics. This comparison will be discussed in appendix \ref{app:comparison}.
Therefore, it seems that trying to find extensive violations of additivity conjectures from the black hole states might be a less fruitful direction than people expected \cite{HP20,Wang22}.

However, our arguments have the potential to give crucial insights into additivity conjectures from an alternative point of view. The reason is that bounding the entanglement of formation $E_F$ plays a significant role in our argument, and there exists a version of additivity conjecture which is profoundly related to $E_F$ \cite{Shor2004}. The entanglement of formation $E_F$ is lower bounded by the entanglement cost $E_C$ \cite{BDSW96}, i.e. $E_F \geq E_C$, in any bipartite systems. Here, the entanglement cost of a bipartite state is defined as the minimal number of Bell pairs one needs to prepare the state with LOCC. It is known that additivity conjectures lead to $E_F = E_C$. Also in this case, while small violations are known \cite{Hastings09,Fukuda2007}, extensive violations have not been found yet \cite{Belinschi2016}. Therefore, it would be interesting to find possible relations between our arguments and this version of additivity conjectures.

\subsection{Atypical microstate counting for black holes}

Although it has long been known that one can find sufficiently many black hole microstates from string theory to match the leading order of the Bekenstein-Hawking entropy since \cite{SV96}, less knowledge is known about the atypical state counting. One of the most successful atypical microstate counting from string theory is that of superstrata. Superstrata \cite{BGRSW15,Shigemori19} are horizonless geometries which are solutions of supergravity and describe microstates of the D1-D5-P black hole. These geometries deviate from the black hole and cap off smoothly. In this sense, the Ryu-Takayanagi surface in it would look very similar to the blue surface shown in the right of figure \ref{fig:BH}. This suggests superstrata are likely to be explicit examples of disentangled states. However, so far, only about $O\left(\exp\left[S_{BH}^{3/4}\right]\right)$ superstrata have been found. This is parametrically smaller than our lower bound of disentangled states which is sufficient to recover the Bekenstein-Hawking entropy. Although not every disentangled state is guaranteed to have a geometrical dual, this comparison is expected to give hints to atypical microstate counting in string theory.

\subsection{Why atypicality is important in studying quantum gravity?}

Let us get some hints from statistical mechanics and argue why studying atypicality is important in understanding quantum gravity. In statistical mechanics, although one knows almost every state in a microscopic theory is typical, it is usually hard to explicitly construct a typical state by experience \cite{SS11}, otherwise, we would just pick up such a typical state to get macroscopic information rather than using canonical ensembles or microcanonical ensembles. 

An intuitive reason is as follows. In order to pick up a microstate from the energy shell in such a complicated many-body system, one usually needs to design a sophisticated algorithm. As a result, the state will be special in the sense that it is picked up by a sophisticated algorithm, and hence tend to be atypical. From this point of view, it is natural why solving supergravity induced from string theory can give so many atypical geometries. This might also be able to help us understand why the chaotic nature of gravity seems to disappear when supersymmetry is involved in AdS$_2$ holography \cite{LMRS22,LMRS22-2}. 

On the other hand, this also suggests that studying generic and collective features of atypical states should be important in understanding the microscopic structure of quantum gravity since the microstates which are easier to get access to are expected to be atypical. 

% \subsection{Operational meanings of correlation measures in holography}

% \subsection{Applications to statistical physics}

\section*{Acknowledgements}
We would like to especially thank Masamichi Miyaji and Jinzhao Wang for careful reading and giving us valuable
comments on a draft of this paper.
We are also grateful to Iosif Bena, Yuya Kusuki, Yasunori Nomura, Hirosi Ooguri, Pratik Rath, Masaki Shigemori, Douglas Stanford, Hiroyasu Tajima, Tadashi Takayanagi and Kotaro Tamaoka for useful discussions. 
ZW was supported by Grant-in-Aid for JSPS Fellows No. 20J23116. ZW was also supported  in part by the National Science Foundation under Grant No. NSF PHY-1748958 and by the Heising-Simons Foundation. YY was supported by Grant-in-Aid for JSPS Fellows No. 20J22982.

\appendix
\section{Axioms for entanglement measures}\label{app:entanglement_measure}

In this appendix, we summarize the axioms (or postulates) \cite{DHR02,PV07,HHH09,BCDS18} that are expected for an entanglement measure $E(\rho)$. We consider a bipartite system $\CH = \CH_A \otimes \CH_B$, and $\rho \in \CD(\CH)$. We assume $\dim \CH_B \leq \dim \CH_A$ without loss of generality. The axioms are as follows.

\begin{description}
  \item[(i)] ~~{If $\rho$ is a separable state, then $E(\rho) = 0$.}  
  \item[(ii)]~{{\it Normalization - }If $\rho$ is a maximally entangled state, then $E(\rho) = \log\dim\CH_B$.} 
  \item[(iii)]{{\it Monotonicity under LOCC - } $E(\rho)$ is monotonically decreasing under LOCC, i.e., for any LOCC operation $\Gamma_{\rm LOCC}(\cdot)$}, 
  \begin{align}
      E(\Gamma_{\rm LOCC}(\rho)) \leq E(\rho).
  \end{align}
  \item[(iv)]{{\it Asymptotic Continuity - } For any sequence of Hilbert space $\{\mathcal{H}_n\}$} and any sequence of states $\{\rho_n\}$ and $\{\sigma_n\}$ with $\rho_n, \sigma_n \in \mathcal{D}(\mathcal{H}_n)$, 
  \begin{align}
      \|\rho_n-\sigma_n\|_1 \rightarrow 0 \quad \Rightarrow \quad \frac{E\left(\rho_{n}\right)-E\left(\sigma_{n}\right)}{1+\log {\rm dim}\mathcal{H}_n} \rightarrow 0.
  \end{align}
  \item[(v)] ~{{\it Additivity - }For any state $\rho \in \CD\left(\CH\right) = \CD\left(\CH_A\otimes \CH_B\right)$, consider its $n$-copy $\rho^{\otimes n} \in \CD\left(\CH^{\otimes n}\right)$. Then for the bipartition $(\mathcal{H}_{A_1}\otimes...\otimes\mathcal{H}_{A_n})\otimes(\mathcal{H}_{B_1}\otimes...\otimes\mathcal{H}_{B_n})$},
  \begin{align}
      E(\rho^{\otimes n}) = nE(\rho).
  \end{align}
  \item[(vi)] ~{\it Convexity - } For any $\rho,\sigma \in \CD(\CH)$ and any $\lambda \in [0,1]$, 
  \begin{align}
      E(\lambda \rho + (1-\lambda) \sigma) \leq \lambda E(\rho)+(1-\lambda) E(\sigma).
  \end{align}
  \item[(vii)] {\it Subadditivity - } For any $\rho \in \CD\left(\CH\right) = \CD\left(\CH_A\otimes \CH_B\right)$ and $\sigma \in \CD\left(\CH'\right) = \CD\left(\CH_{A'}\otimes \CH_{B'}\right)$, consider divide $\CH\otimes\CH'$ into $\left(\CH_A\otimes\CH_{A'}\right) \otimes \left(\CH_B\otimes\CH_{B'}\right)$, then
  \begin{align}
      E(\rho \otimes \sigma) \leq E(\rho) + E(\sigma).
  \end{align}
\end{description}

\section{Alternative proofs of corollary \ref{coro:hol_final} and \ref{coro:qmb}}\label{app:proof}

In this appendix, we present alternative proofs of corollary \ref{coro:hol_final} and \ref{coro:qmb}. These proofs avoid using the cumbersome theorem \ref{thm:Increment_bound}, and therefore turn out to be more simple if one merely wants to show the existence of exponentially many disentangled states or  area law states and does not care about how to construct them. Since the proofs for both cases are similar, we will only present the proof for corollary \ref{coro:qmb}. We will explain how corollary \ref{coro:hol_final} can be proved in a similar way at the end of this appendix.

We firstly show the following theorem which bounds the change of the reflected entropy of microcanonical ensembles under projections. Note that a more general version of this theorem has already been shown in \cite{AR19}.

\begin{customthm}{8} \label{thm:Increment_bound_RE}
  Consider an arbitrary $W$-dimensional subspace $\mathcal{E} \cong \mathbb{C}^W$
  and a density matrix $\rho$ which can be written as $\rho \propto \Pi_\mathcal{E}$,
  where $\Pi_\mathcal{E}$ is the projection operator onto $\mathcal{E}$.
  Then, for any $k$ $(\leq W/4e^2)$ and $(W-k)$-dimensional subspace $\mathcal{E}' \cong \mathbb{C}^{W-k}$ of $\mathcal{E}$,
  the reflected entropy of $\rho' \propto \Pi_{\mathcal{E}'}$
  between $A$ and $B$ can be bounded from two sides as
  \begin{align}
    E_{RE}(\rho') &\leq E_{RE}(\rho) + \sqrt{\frac{k}{W}} \left[\log \frac{W}{k} + 4 \log\dim\mathcal{H}_{B} - 2 \log 2\right], \label{eq:RE_bound}\\
    E_{RE}(\rho') &\geq E_{RE}(\rho) - \sqrt{\frac{k}{W}} \left[\log \frac{W}{k} + 4 \log\dim\mathcal{H}_{B} - 2 \log 2\right],
  \end{align}
  where $\Pi_{\mathcal{E}'}$ are the projection operators onto $\mathcal{E}'$.
\end{customthm}

\begin{proof}
Take an orthogonal basis $\{|\varphi_\nu\rangle\}_{\nu=1,\cdots,W}$ of $\mathcal{E}$ such that
\begin{align}
  \mathcal{E}' &= \mathrm{span} \{|\varphi_\nu\rangle\}_{\nu=k+1,\cdots,W}.
\end{align}
Then we have
\begin{align}
  \rho &= \frac{1}{W} \sum_{\nu=1}^W |\varphi_\nu\rangle\langle\varphi_\nu|,\\
  \rho' &= \frac{1}{W-k} \sum_{\nu=k+1}^W |\varphi_\nu\rangle\langle\varphi_\nu|.
\end{align}
Therefore, according to the definition of the canonical purification in Eq.~\eqref{eq:can_purification},
we have
\begin{align}
  |\sqrt{\rho}\rangle
  &= \frac{1}{\sqrt{W}} \sum_{\nu=1}^W |\varphi_\nu\rangle |\varphi_\nu^*\rangle,\\
  |\sqrt{\rho'}\rangle
  &= \frac{1}{\sqrt{W-k}} \sum_{\nu=k+1}^{W} |\varphi_\nu\rangle |\varphi_\nu^*\rangle.
\end{align}
Hence,
\begin{align}
  \langle\sqrt{\rho'}|\sqrt{\rho}\rangle = \sqrt{1-\frac{k}{W}}.
\end{align}
Since
\begin{align}
  &\|\mathrm{Tr}_{AA^*}|\sqrt{\rho'}\rangle\langle\sqrt{\rho'}|-\mathrm{Tr}_{AA^*}|\sqrt{\rho}\rangle\langle\sqrt{\rho}|\|_1 \nonumber\\
  \leq& \||\sqrt{\rho'}\rangle\langle\sqrt{\rho'}|-|\sqrt{\rho}\rangle\langle\sqrt{\rho}|\|_1 \nonumber\\
  =& 2 \sqrt{1-|\langle\sqrt{\rho'}|\sqrt{\rho}\rangle|^2} \nonumber\\
  =& \sqrt{\frac{4k}{W}},
\end{align}
applying Fannes' inequality, we obtain
\begin{align}
  |E_{RE}(\rho')-E_{RE}(\rho)|
  &= |S(\mathrm{Tr}_{AA^*}|\sqrt{\rho'}\rangle\langle\sqrt{\rho'}|)
  - S(\mathrm{Tr}_{AA^*}|\sqrt{\rho}\rangle\langle\sqrt{\rho}|)| \nonumber\\
  &\leq \sqrt{\frac{4k}{W}} \log\dim\mathcal{H}_{BB*} - \sqrt{\frac{4k}{W}} \log \sqrt{\frac{4k}{W}} \nonumber\\
  &= \sqrt{\frac{k}{W}} \left[\log \frac{W}{k} + 4 \log\dim\mathcal{H}_{B} - 2 \log 2\right].
\end{align}
\end{proof}

Let us then combine conjecture \ref{conjecture} and theorem \ref{thm:Increment_bound_RE}.
As $\mathcal{E}$, let us consider the microcanonical subspace 
\begin{align}
    \mathcal{H}_{[E-\Delta E, E]}
    = \mathrm{span} \left\{ |E_i\rangle \in \mathcal{H} \middle| E_i \in [E-\Delta E, E] \right\}.
\end{align}
Then $\rho$ is the associated microcanonical ensemble $\rho^\mathrm{(micro)}$.
According to the Boltzmann formula, $W \equiv e^{S_{\rm micro}} = O(L^d)$.
By using conjecture \ref{conjecture} and $E_F \leq E_{RE}$,
we can say that there exists at least one area law state $|\varphi_1\rangle$ in $\mathcal{H}_{[E-\Delta E, E]}$.

Then we exclude $|\varphi_1\rangle$ from $\mathcal{H}_{[E-\Delta E, E]}$
and consider $\mathcal{H}_{[E-\Delta E, E]} \cap \left(\mathrm{span}\{|\varphi_1\rangle\}\right)^\perp$ as a new subspace $\mathcal{E}'$.
In this case, since $\log\dim\mathcal{H}_B = O(L^d) = O(S_{\rm micro})$,
according to the upper bound \eqref{eq:RE_bound} given in theorem \ref{thm:Increment_bound_RE},
the reflected entropy of $\rho' \propto \Pi_{\mathcal{E}'}$
is changed by at most $O(e^{-S_{\rm micro}/2} S_{\rm micro})$ from that of $\rho$.
Again, using $E_F \leq E_{RE}$,
we can say that there exists at least one area law state $|\varphi_2\rangle$ in $\mathcal{H}_{[E-\Delta E, E]} \cap \left(\mathrm{span}\{|\varphi_1\rangle\}\right)^\perp$.

According to the bound given in theorem \ref{thm:Increment_bound_RE}, one can at least repeat this procedure for $\exp\left[S_{\rm micro} - O(\log S_{\rm micro})\right]$ times. As a result, one can obtain $\exp\left[S_{\rm micro} - O(\log S_{\rm micro})\right]$ area law states.
By construction, these states are orthogonal. As a result, one can get corollary \ref{coro:qmb}.

The key point of theorem \ref{thm:Increment_bound_RE} is that the change of the reflected entropy when excluding a state from a microcanonical ensemble is sufficiently small. Therefore, if one can show that the change of the entanglement of purification when excluding a state from a microcanonical ensemble is also sufficiently small, the corollary \ref{coro:hol_final} can be shown in the same manner. In fact, such a statement follows directly from the continuity of the entanglement of purification shown in \cite{THLD02,AR19}. Alternatively, one can also use the continuity of the entanglement of formation shown in \cite{Nielsen00} to show corollary \ref{coro:hol_final} and \ref{coro:qmb}.

\section{Comparison of the microcanonical entropy and the canonical entropy} \label{app:comparison}
In this paper, we have utilized the microcanonical entropy $S_\mathrm{micro}$ to count the number of disentangled microstates. 
On the other hand, in Ref.~\cite{HP20}, the canonical entropy $S_\mathrm{can}$, i.e. the von Neumann entropy of the canonical ensemble, was employed for the same purpose. In this appendix, we investigate the relationship between $S_\mathrm{micro}$ and $S_\mathrm{can}$. In the following, we consider a general quantum system with degrees of freedom proportional to $N$.
% Accordingly, the dimension of the Hilbert space is proportional to $e^{\Theta(N)}$.

In comparing these two ensembles, it is important to appropriately choose the microcanonical shell $[E - \Delta E, E]$ with respect to the inverse temperature $\beta$ of the canonical ensemble. 
Here, in accordance with Ref.~\cite{HP20}, we take the microcanonical shell to be centered at the saddle point $E_\mathrm{can}^*$ of the energy distribution in the canonical ensemble and to have a width of the same order of magnitude as the energy fluctuations in the canonical ensemble. This corresponds to setting $E$ and $\Delta E$ as $E = E_\mathrm{can}^* + \Delta E /2$ and $\Delta E = \Theta (N^{1/2})$ \footnote{We write $f(N)=\Theta(g(N))$ when $f$ is bounded both above and below by $g$ asymptotically as $N\to\infty$.}.

To discuss the asymptotic behavior in the large degree of freedom $N$ limit, we introduce the energy per degree of freedom, $u \equiv U/N$.
% In a homogeneous state, $u$ is just the energy density.
We then write $\epsilon \equiv E/N$, $\delta\epsilon \equiv \Delta E/N$, and $\epsilon_\mathrm{can}^* \equiv E_\mathrm{can}^*/N$. In this notation, the condition for the microcanonical shell can be rephrased as follows:
\begin{align}
  \epsilon &= \epsilon_\mathrm{can}^* + \delta\epsilon/2, \label{eq:shell_center}\\
  \delta\epsilon &= \Theta (N^{-1/2}). \label{eq:shell_width}
\end{align}

Let $\Gamma(u)$ represent the density of microstates at energy per degree of freedom $u$. We assume that the system is consistent with thermodynamics in the following sense: $\sigma(u) \equiv \frac{1}{N} \log \Gamma(u)$ converges to an $N$-independent concave function (namely the thermodynamic entropy per degree of freedom) in the large degree of freedom limit $N\to\infty$. For simplicity of discussion, we also assume that the inverse temperature $\beta$ is positive and does not scale with $N$ \footnote{The following argument can be generalized straightforwardly to the case where the inverse temperature is negative.}. That is, we assume that
\begin{align}
  \frac{d\sigma}{du}(\epsilon) = \Theta(N^0) >0. \label{eq:positive-temperature}
\end{align}

First, we evaluate the microcanonical entropy $S_\mathrm{micro}$.
Since $\sigma(u)$ is concave for sufficiently large $N$, for any $u \in [\epsilon - \delta\epsilon, \epsilon]$, it holds that
\begin{align}
  \sigma(u)
  \geq \frac{\sigma(\epsilon) - \sigma(\epsilon - \delta\epsilon)}{\delta\epsilon} u
  + \frac{\epsilon \sigma(\epsilon - \delta\epsilon) - (\epsilon - \delta\epsilon) \sigma(\epsilon)}{\delta\epsilon}
  \equiv \varsigma(u).
\end{align}
Using this, we find
\begin{align}
  \dim \CH_{[E-\Delta E,E]}
  % = \sum_{E_i \in [E-\Delta E, E]} 1
  &= \int_{\epsilon-\delta\epsilon}^{\epsilon} du\ \Gamma(u)
  = \int_{\epsilon-\delta\epsilon}^{\epsilon} du\ e^{N\sigma(u)}
  \geq \int_{\epsilon-\delta\epsilon}^{\epsilon} du\ e^{N\varsigma(u)} \nonumber\\
  &= e^{N\sigma(\epsilon)}
  \times \frac{\delta\epsilon}{N[\sigma(\epsilon)-\sigma(\epsilon-\delta\epsilon)]}
  \times \left( 1 - e^{-N[\sigma(\epsilon)-\sigma(\epsilon - \delta\epsilon)]} \right). \label{eq:W_lower-bound}
\end{align}
Expanding $\sigma(u)$ around $\epsilon$, we find
\begin{align}
  \sigma(\epsilon - \delta\epsilon)
  = \sigma(\epsilon) - \frac{d\sigma}{du}(\epsilon) \delta\epsilon + O({\delta\epsilon}^2).
\end{align}
Thus, using Eqs.~\eqref{eq:shell_width} and \eqref{eq:positive-temperature}, we have
\begin{align}
  \sigma(\epsilon) - \sigma(\epsilon - \delta\epsilon)
  = \Theta(\delta\epsilon)
  = \Theta(N^{-1/2}).
\end{align}
Substituting this into Eq.~\eqref{eq:W_lower-bound}, we have
\begin{align}
  S_\mathrm{micro}
  = \log \dim \CH_{[E-\Delta E,E]}
  \geq N\sigma(\epsilon) - \log N + \Theta(N^0).\label{eq:microcanonical_entropy}
\end{align}

Next, we evaluate the canonical entropy:
\begin{align}
  S_\mathrm{can} = \log Z_\mathrm{can} + \beta E_\mathrm{can}.
\end{align}
Here, $Z_\mathrm{can} = \mathrm{Tr} \left[ e^{- \beta H} \right]$ is the partition function and $E_\mathrm{can} = \mathrm{Tr} \left[ H e^{- \beta H} \right] \left/ \mathrm{Tr} \left[ e^{- \beta H} \right] \right.$ is the energy function.
Then $Z_\mathrm{can}$ can be expressed as
\begin{align}
  Z_\mathrm{can}
  = \int du\ \Gamma(u) e^{- N \beta u}
  = \int du\ e^{N[\sigma(u) - \beta u]}.
\end{align}
Evaluating this integral by Laplace's method, we find \cite{YS19}
\begin{align}
  \log Z_\mathrm{can} = N [\sigma(\epsilon_\mathrm{can}^*) - \beta \epsilon_\mathrm{can}^*] + O(\log N).
\end{align}
In the same way, it can be shown that
\begin{align}
  \epsilon_\mathrm{can} = \epsilon_\mathrm{can}^* + O(N^{-1}).
\end{align}
Therefore, we obtain
\begin{align}
  S_\mathrm{can} = N\sigma(\epsilon_\mathrm{can}^*) + O(\log N). \label{eq:canonical_entropy}
\end{align}

Finally, we investigate the relationship between $S_\mathrm{micro}$ and $S_\mathrm{can}$. Using Eqs.~\eqref{eq:microcanonical_entropy} and \eqref{eq:canonical_entropy}, we have
\begin{align}
  S_\mathrm{micro} - S_\mathrm{can}
  \geq N\sigma(\epsilon) - N\sigma(\epsilon_\mathrm{can}^*) + O(\log N)
  = \Theta (N^{1/2}).
\end{align}
In the last equality, we have used Eqs.~\eqref{eq:shell_center}-\eqref{eq:positive-temperature}. Therefore, for sufficiently large $N$, it holds that\footnote{Readers who know the principle of maximum entropy may get confused about this result, since it is often said ``the canonical ensemble gives a larger entropy than any other ensembles". However, we would like to note that a prerequisite of this statement is that all the ensembles considered give exactly the same energy expectation value. In the case we have discussed, however, the energy expectation value computed from the microcanonical ensemble is larger than that computed from the canonical ensemble at the $\Theta(N^{1/2})$ order.}
\begin{align}
  S_\mathrm{micro} \geq S_\mathrm{can} + \Theta(N^{1/2}).
\end{align}

In the main text of this paper, we have shown that for any microcanonical ensemble as long as the semiclassical gravity dual holds, one can find at least $\exp\left[S_{\rm micro} - f(S_{\rm micro})\right]$ orthogonal disentangled states for any $f(x)\in \omega (1)$ in the corresponding microcanonical sub-Hilbert space. On the other hand, it is shown in \cite{HP20} that, assuming there is no extensive violations of additivity conjectures, one can find at least $\exp\left[S_{\rm can} - O(1))\right]$ orthogonal disentangled states from a microcanonical sub-Hilbert space whose energy window is $[E_{\rm can}^* - \Delta E/2, E_{\rm can}^* + \Delta E/2]$ where $\Delta E = \Theta(N^{1/2})$.

From the results above, it is straightforward to see that there exists $f(x) \in \omega(1)$ such that $\exp\left[S_{\rm micro} - f(S_{\rm micro})\right] > \exp\left[S_{\rm can} - O(1)\right]$. This implies that our lower bound surpasses the one obtained in Ref.~\cite{HP20}. In other words, our bound is stronger. It is noteworthy that in contrast to Ref.~\cite{HP20}, our results do not rely on the assumption of the absence of extensive violations of additivity conjectures and are therefore more robust in this respect.

\newpage
\bibliographystyle{jhep}
\bibliography{Microstates}

\providecommand{\href}[2]{#2}\begingroup\raggedright\begin{thebibliography}{10}

\bibitem{Goldstein2006}
S.~Goldstein, J.~L. Lebowitz, R.~Tumulka and N.~Zangh{\`{i}}, \emph{{Canonical
  typicality}},
  \href{https://doi.org/10.1103/PhysRevLett.96.050403}{\emph{Phys. Rev. Lett.}
  {\bfseries 96} (2006) 050403}
  [\href{https://arxiv.org/abs/0511091}{{\ttfamily 0511091}}].

\bibitem{Popescu2006}
S.~Popescu, A.~J. Short and A.~Winter, \emph{{Entanglement and the foundations
  of statistical mechanics}},
  \href{https://doi.org/10.1038/nphys444}{\emph{Nat. Phys.} {\bfseries 2}
  (2006) 754} [\href{https://arxiv.org/abs/0511225}{{\ttfamily 0511225}}].

\bibitem{Reimann2007}
P.~Reimann, \emph{{Typicality for generalized microcanonical ensemble}},
  \href{https://doi.org/10.1103/PhysRevLett.99.160404}{\emph{Phys. Rev. Lett.}
  {\bfseries 99} (2007) 160404}
  [\href{https://arxiv.org/abs/0710.4214}{{\ttfamily 0710.4214}}].

\bibitem{Muller2015}
M.~P. M{\"{u}}ller, E.~Adlam, L.~Masanes and N.~Wiebe, \emph{{Thermalization
  and Canonical Typicality in Translation-Invariant Quantum Lattice Systems}},
  \href{https://doi.org/10.1007/s00220-015-2473-y}{\emph{Commun. Math. Phys.}
  {\bfseries 340} (2015) 499}
  [\href{https://arxiv.org/abs/1312.7420}{{\ttfamily 1312.7420}}].

\bibitem{Mori2018}
T.~Mori, T.~Ikeda, E.~Kaminishi and M.~Ueda, \emph{{Thermalization and
  prethermalization in isolated quantum systems: a theoretical overview}},
  \href{https://doi.org/10.1088/1361-6455/aabcdf}{\emph{J. Phys. B} {\bfseries
  51} (2018) 112001} [\href{https://arxiv.org/abs/1712.08790}{{\ttfamily
  1712.08790}}].

\bibitem{Hawking75}
S.~W. Hawking, \emph{{Particle Creation by Black Holes}},
  \href{https://doi.org/10.1007/BF02345020}{\emph{Commun. Math. Phys.}
  {\bfseries 43} (1975) 199}.

\bibitem{Page93}
D.~N. Page, \emph{{Information in black hole radiation}},
  \href{https://doi.org/10.1103/PhysRevLett.71.3743}{\emph{Phys. Rev. Lett.}
  {\bfseries 71} (1993) 3743}
  [\href{https://arxiv.org/abs/hep-th/9306083}{{\ttfamily hep-th/9306083}}].

\bibitem{AMPS12}
A.~Almheiri, D.~Marolf, J.~Polchinski and J.~Sully, \emph{{Black Holes:
  Complementarity or Firewalls?}},
  \href{https://doi.org/10.1007/JHEP02(2013)062}{\emph{JHEP} {\bfseries 02}
  (2013) 062} [\href{https://arxiv.org/abs/1207.3123}{{\ttfamily 1207.3123}}].

\bibitem{AMPSS13}
A.~Almheiri, D.~Marolf, J.~Polchinski, D.~Stanford and J.~Sully, \emph{{An
  Apologia for Firewalls}},
  \href{https://doi.org/10.1007/JHEP09(2013)018}{\emph{JHEP} {\bfseries 09}
  (2013) 018} [\href{https://arxiv.org/abs/1304.6483}{{\ttfamily 1304.6483}}].

\bibitem{Mathur05}
S.~D. Mathur, \emph{{The Fuzzball proposal for black holes: An Elementary
  review}}, \href{https://doi.org/10.1002/prop.200410203}{\emph{Fortsch. Phys.}
  {\bfseries 53} (2005) 793}
  [\href{https://arxiv.org/abs/hep-th/0502050}{{\ttfamily hep-th/0502050}}].

\bibitem{NVW12}
Y.~Nomura, J.~Varela and S.~J. Weinberg, \emph{{Black Holes, Information, and
  Hilbert Space for Quantum Gravity}},
  \href{https://doi.org/10.1103/PhysRevD.87.084050}{\emph{Phys. Rev. D}
  {\bfseries 87} (2013) 084050}
  [\href{https://arxiv.org/abs/1210.6348}{{\ttfamily 1210.6348}}].

\bibitem{HM13}
T.~Hartman and J.~Maldacena, \emph{{Time Evolution of Entanglement Entropy from
  Black Hole Interiors}},
  \href{https://doi.org/10.1007/JHEP05(2013)014}{\emph{JHEP} {\bfseries 05}
  (2013) 014} [\href{https://arxiv.org/abs/1303.1080}{{\ttfamily 1303.1080}}].

\bibitem{HP20}
P.~Hayden and G.~Penington, \emph{{Black hole microstates vs. the additivity
  conjectures}},  \href{https://arxiv.org/abs/2012.07861}{{\ttfamily
  2012.07861}}.

\bibitem{Maldacena97}
J.~M. Maldacena, \emph{{The Large N limit of superconformal field theories and
  supergravity}}, \href{https://doi.org/10.1023/A:1026654312961,
  10.4310/ATMP.1998.v2.n2.a1}{\emph{Int. J. Theor. Phys.} {\bfseries 38} (1999)
  1113} [\href{https://arxiv.org/abs/hep-th/9711200}{{\ttfamily
  hep-th/9711200}}].

\bibitem{RT06}
S.~Ryu and T.~Takayanagi, \emph{{Holographic derivation of entanglement entropy
  from AdS/CFT}},
  \href{https://doi.org/10.1103/PhysRevLett.96.181602}{\emph{Phys. Rev. Lett.}
  {\bfseries 96} (2006) 181602}
  [\href{https://arxiv.org/abs/hep-th/0603001}{{\ttfamily hep-th/0603001}}].

\bibitem{RT06b}
S.~Ryu and T.~Takayanagi, \emph{{Aspects of Holographic Entanglement Entropy}},
  \href{https://doi.org/10.1088/1126-6708/2006/08/045}{\emph{JHEP} {\bfseries
  08} (2006) 045} [\href{https://arxiv.org/abs/hep-th/0605073}{{\ttfamily
  hep-th/0605073}}].

\bibitem{HRT07}
V.~E. Hubeny, M.~Rangamani and T.~Takayanagi, \emph{{A Covariant holographic
  entanglement entropy proposal}},
  \href{https://doi.org/10.1088/1126-6708/2007/07/062}{\emph{JHEP} {\bfseries
  07} (2007) 062} [\href{https://arxiv.org/abs/0705.0016}{{\ttfamily
  0705.0016}}].

\bibitem{Wall12}
A.~C. Wall, \emph{{Maximin Surfaces, and the Strong Subadditivity of the
  Covariant Holographic Entanglement Entropy}},
  \href{https://doi.org/10.1088/0264-9381/31/22/225007}{\emph{Class. Quant.
  Grav.} {\bfseries 31} (2014) 225007}
  [\href{https://arxiv.org/abs/1211.3494}{{\ttfamily 1211.3494}}].

\bibitem{FLM13}
T.~Faulkner, A.~Lewkowycz and J.~Maldacena, \emph{{Quantum corrections to
  holographic entanglement entropy}},
  \href{https://doi.org/10.1007/JHEP11(2013)074}{\emph{JHEP} {\bfseries 11}
  (2013) 074} [\href{https://arxiv.org/abs/1307.2892}{{\ttfamily 1307.2892}}].

\bibitem{EW15}
N.~Engelhardt and A.~C. Wall, \emph{{Quantum Extremal Surfaces: Holographic
  Entanglement Entropy beyond the Classical Regime}},
  \href{https://doi.org/10.1007/JHEP01(2015)073}{\emph{JHEP} {\bfseries 01}
  (2015) 073} [\href{https://arxiv.org/abs/1408.3203}{{\ttfamily 1408.3203}}].

\bibitem{AP20}
C.~Akers and G.~Penington, \emph{{Leading order corrections to the quantum
  extremal surface prescription}},
  \href{https://doi.org/10.1007/JHEP04(2021)062}{\emph{JHEP} {\bfseries 04}
  (2021) 062} [\href{https://arxiv.org/abs/2008.03319}{{\ttfamily
  2008.03319}}].

\bibitem{Shor2004}
P.~W. Shor, \emph{{Equivalence of additivity questions in quantum information
  theory}}, \href{https://doi.org/10.1007/s00220-003-0981-7}{\emph{Commun.
  Math. Phys.} {\bfseries 246} (2004) 453}
  [\href{https://arxiv.org/abs/0305035}{{\ttfamily 0305035}}].

\bibitem{Hastings09}
M.~B. Hastings, \emph{Superadditivity of communication capacity using entangled
  inputs}, \href{https://doi.org/10.1038/nphys1224}{\emph{Nature Physics}
  {\bfseries 5} (2009) 255–257}.

\bibitem{JAVWY20}
Z.~Ji, A.~Natarajan, T.~Vidick, J.~Wright and H.~Yuen, \emph{{MIP*=RE}},
  \href{https://arxiv.org/abs/2001.04383}{{\ttfamily 2001.04383}}.

\bibitem{Wang22}
J.~Wang, \emph{{Beyond islands: a free probabilistic approach}},
  \href{https://doi.org/10.1007/JHEP10(2023)040}{\emph{JHEP} {\bfseries 10}
  (2023) 040} [\href{https://arxiv.org/abs/2209.10546}{{\ttfamily
  2209.10546}}].

\bibitem{UT17}
T.~Takayanagi and K.~Umemoto, \emph{{Entanglement of purification through
  holographic duality}},
  \href{https://doi.org/10.1038/s41567-018-0075-2}{\emph{Nature Phys.}
  {\bfseries 14} (2018) 573}
  [\href{https://arxiv.org/abs/1708.09393}{{\ttfamily 1708.09393}}].

\bibitem{NDHZS17}
P.~Nguyen, T.~Devakul, M.~G. Halbasch, M.~P. Zaletel and B.~Swingle,
  \emph{{Entanglement of purification: from spin chains to holography}},
  \href{https://doi.org/10.1007/JHEP01(2018)098}{\emph{JHEP} {\bfseries 01}
  (2018) 098} [\href{https://arxiv.org/abs/1709.07424}{{\ttfamily
  1709.07424}}].

\bibitem{THLD02}
B.~M. Terhal, M.~Horodecki, D.~W. Leung and D.~P. DiVincenzo, \emph{The
  entanglement of purification},
  \href{https://doi.org/10.1063/1.1498001}{\emph{Journal of Mathematical
  Physics} {\bfseries 43} (2002) 4286–4298}
  [\href{https://arxiv.org/abs/quant-ph/0202044}{{\ttfamily
  quant-ph/0202044}}].

\bibitem{BDSW96}
C.~H. Bennett, D.~P. DiVincenzo, J.~A. Smolin and W.~K. Wootters,
  \emph{Mixed-state entanglement and quantum error correction},
  \href{https://doi.org/10.1103/physreva.54.3824}{\emph{Physical Review A}
  {\bfseries 54} (1996) 3824–3851}
  [\href{https://arxiv.org/abs/quant-ph/9604024}{{\ttfamily
  quant-ph/9604024}}].

\bibitem{MS19}
C.~Murthy and M.~Srednicki, \emph{{Structure of chaotic eigenstates and their
  entanglement entropy}},
  \href{https://doi.org/10.1103/PhysRevE.100.022131}{\emph{Phys. Rev. E}
  {\bfseries 100} (2019) 022131}
  [\href{https://arxiv.org/abs/1906.04295}{{\ttfamily 1906.04295}}].

\bibitem{Srednicki94}
M.~Srednicki, \emph{Chaos and quantum thermalization},
  \href{https://doi.org/10.1103/PhysRevE.50.888}{\emph{Phys. Rev. E} {\bfseries
  50} (1994) 888} [\href{https://arxiv.org/abs/cond-mat/9403051}{{\ttfamily
  cond-mat/9403051}}].

\bibitem{MWW20}
D.~Marolf, S.~Wang and Z.~Wang, \emph{{Probing phase transitions of holographic
  entanglement entropy with fixed area states}},
  \href{https://doi.org/10.1007/JHEP12(2020)084}{\emph{JHEP} {\bfseries 12}
  (2020) 084} [\href{https://arxiv.org/abs/2006.10089}{{\ttfamily
  2006.10089}}].

\bibitem{DW20}
X.~Dong and H.~Wang, \emph{{Enhanced corrections near holographic entanglement
  transitions: a chaotic case study}},
  \href{https://doi.org/10.1007/JHEP11(2020)007}{\emph{JHEP} {\bfseries 11}
  (2020) 007} [\href{https://arxiv.org/abs/2006.10051}{{\ttfamily
  2006.10051}}].

\bibitem{DF19}
S.~Dutta and T.~Faulkner, \emph{{A canonical purification for the entanglement
  wedge cross-section}},
  \href{https://doi.org/10.1007/JHEP03(2021)178}{\emph{JHEP} {\bfseries 03}
  (2021) 178} [\href{https://arxiv.org/abs/1905.00577}{{\ttfamily
  1905.00577}}].

\bibitem{DHR02}
M.~J. Donald, M.~Horodecki and O.~Rudolph, \emph{The uniqueness theorem for
  entanglement measures},
  \href{https://doi.org/10.1063/1.1495917}{\emph{Journal of Mathematical
  Physics} {\bfseries 43} (2002) 4252–4272}
  [\href{https://arxiv.org/abs/quant-ph/0105017}{{\ttfamily
  quant-ph/0105017}}].

\bibitem{PV07}
M.~B. Plenio and S.~Virmani, \emph{{An Introduction to entanglement measures}},
  {\emph{Quant. Inf. Comput.} {\bfseries 7} (2007) 1}
  [\href{https://arxiv.org/abs/quant-ph/0504163}{{\ttfamily
  quant-ph/0504163}}].

\bibitem{HHH09}
R.~Horodecki, P.~Horodecki, M.~Horodecki and K.~Horodecki, \emph{{Quantum
  entanglement}}, \href{https://doi.org/10.1103/RevModPhys.81.865}{\emph{Rev.
  Mod. Phys.} {\bfseries 81} (2009) 865}
  [\href{https://arxiv.org/abs/quant-ph/0702225}{{\ttfamily
  quant-ph/0702225}}].

\bibitem{BCDS18}
G.~Benenti, G.~Casati, D.~Rossini and G.~Strini, \emph{Principles of Quantum
  Computation and Information}. WORLD SCIENTIFIC, 2018,
  \href{https://doi.org/10.1142/10909}{10.1142/10909}.

\bibitem{CKNvR12}
B.~Czech, J.~L. Karczmarek, F.~Nogueira and M.~Van~Raamsdonk, \emph{{The
  Gravity Dual of a Density Matrix}},
  \href{https://doi.org/10.1088/0264-9381/29/15/155009}{\emph{Class. Quant.
  Grav.} {\bfseries 29} (2012) 155009}
  [\href{https://arxiv.org/abs/1204.1330}{{\ttfamily 1204.1330}}].

\bibitem{HHLR14}
M.~Headrick, V.~E. Hubeny, A.~Lawrence and M.~Rangamani, \emph{{Causality \&
  holographic entanglement entropy}},
  \href{https://doi.org/10.1007/JHEP12(2014)162}{\emph{JHEP} {\bfseries 12}
  (2014) 162} [\href{https://arxiv.org/abs/1408.6300}{{\ttfamily 1408.6300}}].

\bibitem{MT15}
M.~Miyaji and T.~Takayanagi, \emph{{Surface/State Correspondence as a
  Generalized Holography}},
  \href{https://doi.org/10.1093/ptep/ptv089}{\emph{PTEP} {\bfseries 2015}
  (2015) 073B03} [\href{https://arxiv.org/abs/1503.03542}{{\ttfamily
  1503.03542}}].

\bibitem{Swingle09}
B.~Swingle, \emph{{Entanglement Renormalization and Holography}},
  \href{https://doi.org/10.1103/PhysRevD.86.065007}{\emph{Phys. Rev. D}
  {\bfseries 86} (2012) 065007}
  [\href{https://arxiv.org/abs/0905.1317}{{\ttfamily 0905.1317}}].

\bibitem{PYHP15}
F.~Pastawski, B.~Yoshida, D.~Harlow and J.~Preskill, \emph{{Holographic quantum
  error-correcting codes: Toy models for the bulk/boundary correspondence}},
  \href{https://doi.org/10.1007/JHEP06(2015)149}{\emph{JHEP} {\bfseries 06}
  (2015) 149} [\href{https://arxiv.org/abs/1503.06237}{{\ttfamily
  1503.06237}}].

\bibitem{HNQTY16}
P.~Hayden, S.~Nezami, X.-L. Qi, N.~Thomas, M.~Walter and Z.~Yang,
  \emph{{Holographic duality from random tensor networks}},
  \href{https://doi.org/10.1007/JHEP11(2016)009}{\emph{JHEP} {\bfseries 11}
  (2016) 009} [\href{https://arxiv.org/abs/1601.01694}{{\ttfamily
  1601.01694}}].

\bibitem{GKP98}
S.~S. Gubser, I.~R. Klebanov and A.~M. Polyakov, \emph{{Gauge theory
  correlators from noncritical string theory}},
  \href{https://doi.org/10.1016/S0370-2693(98)00377-3}{\emph{Phys. Lett.}
  {\bfseries B428} (1998) 105}
  [\href{https://arxiv.org/abs/hep-th/9802109}{{\ttfamily hep-th/9802109}}].

\bibitem{Witten98}
E.~Witten, \emph{{Anti-de Sitter space and holography}},
  \href{https://doi.org/10.4310/ATMP.1998.v2.n2.a2}{\emph{Adv. Theor. Math.
  Phys.} {\bfseries 2} (1998) 253}
  [\href{https://arxiv.org/abs/hep-th/9802150}{{\ttfamily hep-th/9802150}}].

\bibitem{Maldacena01}
J.~M. Maldacena, \emph{{Eternal black holes in anti-de Sitter}},
  \href{https://doi.org/10.1088/1126-6708/2003/04/021}{\emph{JHEP} {\bfseries
  04} (2003) 021} [\href{https://arxiv.org/abs/hep-th/0106112}{{\ttfamily
  hep-th/0106112}}].

\bibitem{Marolf18}
D.~Marolf, \emph{{Microcanonical Path Integrals and the Holography of small
  Black Hole Interiors}},
  \href{https://doi.org/10.1007/JHEP09(2018)114}{\emph{JHEP} {\bfseries 09}
  (2018) 114} [\href{https://arxiv.org/abs/1808.00394}{{\ttfamily
  1808.00394}}].

\bibitem{Barthel2017}
T.~Barthel, \emph{{One-dimensional quantum systems at finite temperatures can
  be simulated efficiently on classical computers}},
  \href{https://arxiv.org/abs/1708.09349}{{\ttfamily 1708.09349}}.

\bibitem{Kuwahara2021}
T.~Kuwahara, {\'{A}}.~M. Alhambra and A.~Anshu, \emph{{Improved Thermal Area
  Law and Quasilinear Time Algorithm for Quantum Gibbs States}},
  \href{https://doi.org/10.1103/PhysRevX.11.011047}{\emph{Phys. Rev. X}
  {\bfseries 11} (2021) 011047}
  [\href{https://arxiv.org/abs/2007.11174}{{\ttfamily 2007.11174}}].

\bibitem{KTWY23}
Y.~Kusuki, K.~Tamaoka, Z.~Wei and Y.~Yoneta, \emph{Efficient simulation of
  low-temperature physics in one-dimensional gapless systems},
  \href{https://doi.org/10.1103/PhysRevB.110.L041122}{\emph{Phys. Rev. B}
  {\bfseries 110} (2024) L041122}
  [\href{https://arxiv.org/abs/2309.02519}{{\ttfamily 2309.02519}}].

\bibitem{Fukuda2007}
M.~Fukuda and M.~M. Wolf, \emph{{Simplifying additivity problems using direct
  sum constructions}}, \href{https://doi.org/10.1063/1.2746128}{\emph{J. Math.
  Phys.} {\bfseries 48} (2007) 072101}.

\bibitem{Belinschi2016}
S.~T. Belinschi, B.~Collins and I.~Nechita, \emph{{Almost one bit violation for
  the additivity of the minimum output entropy}},
  \href{https://doi.org/10.1007/s00220-015-2561-z}{\emph{Commun. Math. Phys.}
  {\bfseries 341} (2016) 885}
  [\href{https://arxiv.org/abs/1305.1567}{{\ttfamily 1305.1567}}].

\bibitem{SV96}
A.~Strominger and C.~Vafa, \emph{{Microscopic origin of the Bekenstein-Hawking
  entropy}}, \href{https://doi.org/10.1016/0370-2693(96)00345-0}{\emph{Phys.
  Lett. B} {\bfseries 379} (1996) 99}
  [\href{https://arxiv.org/abs/hep-th/9601029}{{\ttfamily hep-th/9601029}}].

\bibitem{BGRSW15}
I.~Bena, S.~Giusto, R.~Russo, M.~Shigemori and N.~P. Warner, \emph{{Habemus
  Superstratum! A constructive proof of the existence of superstrata}},
  \href{https://doi.org/10.1007/JHEP05(2015)110}{\emph{JHEP} {\bfseries 05}
  (2015) 110} [\href{https://arxiv.org/abs/1503.01463}{{\ttfamily
  1503.01463}}].

\bibitem{Shigemori19}
M.~Shigemori, \emph{{Counting Superstrata}},
  \href{https://doi.org/10.1007/JHEP10(2019)017}{\emph{JHEP} {\bfseries 10}
  (2019) 017} [\href{https://arxiv.org/abs/1907.03878}{{\ttfamily
  1907.03878}}].

\bibitem{SS11}
S.~Sugiura and A.~Shimizu, \emph{{Thermal Pure Quantum States at Finite
  Temperature}},
  \href{https://doi.org/10.1103/PhysRevLett.108.240401}{\emph{Phys. Rev. Lett.}
  {\bfseries 108} (2012) 240401}
  [\href{https://arxiv.org/abs/1112.0740}{{\ttfamily 1112.0740}}].

\bibitem{LMRS22}
H.~W. Lin, J.~Maldacena, L.~Rozenberg and J.~Shan, \emph{{Holography for people
  with no time}},
  \href{https://doi.org/10.21468/SciPostPhys.14.6.150}{\emph{SciPost Phys.}
  {\bfseries 14} (2023) 150}
  [\href{https://arxiv.org/abs/2207.00407}{{\ttfamily 2207.00407}}].

\bibitem{LMRS22-2}
H.~W. Lin, J.~Maldacena, L.~Rozenberg and J.~Shan, \emph{{Looking at
  supersymmetric black holes for a very long time}},
  \href{https://doi.org/10.21468/SciPostPhys.14.5.128}{\emph{SciPost Phys.}
  {\bfseries 14} (2023) 128}
  [\href{https://arxiv.org/abs/2207.00408}{{\ttfamily 2207.00408}}].

\bibitem{AR19}
C.~Akers and P.~Rath, \emph{{Entanglement Wedge Cross Sections Require
  Tripartite Entanglement}},
  \href{https://doi.org/10.1007/JHEP04(2020)208}{\emph{JHEP} {\bfseries 04}
  (2020) 208} [\href{https://arxiv.org/abs/1911.07852}{{\ttfamily
  1911.07852}}].

\bibitem{Nielsen00}
M.~A. Nielsen, \emph{Continuity bounds for entanglement},
  \href{https://doi.org/10.1103/physreva.61.064301}{\emph{Physical Review A}
  {\bfseries 61} (2000) }
  [\href{https://arxiv.org/abs/quant-ph/9908086}{{\ttfamily
  quant-ph/9908086}}].

\bibitem{YS19}
Y.~Yoneta and A.~Shimizu, \emph{{Squeezed ensemble for systems with first-order
  phase transitions}},
  \href{https://doi.org/10.1103/PhysRevB.99.144105}{\emph{Phys. Rev. B}
  {\bfseries 99} (2019) 144105}
  [\href{https://arxiv.org/abs/1903.04111}{{\ttfamily 1903.04111}}].

\end{thebibliography}\endgroup

\end{document}